\documentstyle[12pt,epsf]{article}
\newcommand{\beq}{\begin{equation}}
\newcommand{\eeq}{\end{equation}}
\newcommand{\beqa}{\begin{eqnarray}}
\newcommand{\eeqa}{\end{eqnarray}}

\newcommand{\Z}{{\bf Z}}

\newcommand{\R}{{\bf R}}
\newcommand{\C}{{\bf C}}

\newcommand{\rank}{{\rm rank}\,}
\newcommand{\NSp}{NS${}^{\prime}$~}
\newcommand{\tilQ}{\widetilde{Q}}
\newcommand{\bk}{{\bf k}}

\newcommand{\CP}{{\bf C}{\rm P}}
\newcommand{\tilm}{\widetilde{m}}
\newcommand{\btilm}{\widehat{m}}
\newcommand{\tilW}{\widetilde{W}}
\newcommand{\tilj}{\widetilde{j}}

\newcommand{\Tr}{{\rm Tr}}
\newcommand{\bQ}{\overline{Q}}
\newcommand{\bD}{\overline{D}}
\newcommand{\blambda}{\overline{\lambda}}
\newcommand{\UA}{U(1)_{\rm A}}
\newcommand{\UV}{U(1)_{\rm V}}
\newcommand{\e}{{\rm e}}
\newcommand{\dd}{{\rm d}}
\newcommand{\eeff}{e_{\it eff}}
\newcommand{\elst}{\ell_{\it st}}
\newcommand{\elel}{\ell_{11}}

\makeatletter
  
  \@addtoreset{equation}{section}
\makeatother

\begin{document}

\begin{titlepage}

\begin{center}
\today
\hfill IASSNS-HEP-97/81\\
\hfill UCB-PTH-97/39\\
\hfill LBNL-40571\\
\hfill hep-th/9707192
\vspace{2cm}

{\Large\bf Branes and $N=2$ Theories in Two Dimensions}

\vspace{2cm}
{\large Amihay Hanany $\!{}^{*}$ and\,\,\,
Kentaro Hori ${}^{\dag,\ddag}$}

\vspace{1cm}

{\it ${}^*$School of Natural Sciences,
Institute for Advanced Study\\
Olden Lane, Princeton, NJ 08540, USA\\[0.3cm]
${}^{\dag}$Department of Physics,
University of California at Berkeley\\
Berkeley, CA 94720, USA\\[0.3cm]
${}^{\ddag}$Physics Division,
Lawrence Berkeley National Laboratory\\
Berkeley, CA 94720, USA}

\end{center}

\vspace{1.5cm}
\begin{abstract}
Type IIA brane configurations are used to construct $N=2$
supersymmetric gauge theories in two dimensions.
Using localization of chiral multiplets in ten-dimensional
spacetime, supersymmetric non-linear sigma models with target space
such as $\CP^{n-1}$ and the Grassmann manifolds are studied in detail.
The quantum properties of these models are realized in $M$ theory
by taking the strong Type IIA coupling limit. The brane picture implies
an equivalence between the parameter space of $N=2$ supersymmetric
theories in two dimensions and the moduli space of vacua of $N=2$
supersymmetric gauge theories in four dimensions. Effects like
level-rank duality are interpreted in the brane picture as continuation
past infinite coupling. The BPS solitons of the $\CP^{n-1}$ model are
identified as topological excitations of a membrane
and their masses are computed.
This provides the brane realization of higher rank tensor
representations of the flavor group.
\end{abstract}

\end{titlepage}

\baselineskip=0.7cm
\parskip=0.05cm

\section{Introduction}

~~~~~The realization of supersymmetric gauge theories
using various branes in string theory,
with the aid of some string theory dualities,
enables us to make
various predictions on the dynamical effects
in the strong coupling regime, as was first exhibited in \cite{HW}.
Phenomena in theories with eight supercharges were studied subsequently
in \cite{DHOOY,WittenM5,AH,karch,barak,lll,tel}.
Theories with four supercharges were constructed in \cite{Kutasov}
and studied further in
\cite{DHOY,OV,Barbon,
jon,AO,BH,ted,Adam,AH,tatar,Ahn,AhnT,HZ,HOO,W97,telQCD,AOT}.
Supersymmetric gauge theories can also be studied from
string theory
by realizing the gauge symmetry as
a singular geometry of the string compactification \cite{SKCV}.
This method has also been developed extensively in various directions.
\footnote{In particular, in \cite{KLMVW} it was noted
that solutions of some $N=2$ theories are realized as the
configurations of Type IIA fivebranes in flat space-time.}

One important aspect of the discussion in theories with
four supercharges is the realization of chiral symmetry
and chiral gauge theory in terms of branes.
A proposal of chiral symmetry realization
was given in \cite{BH} in which it was also proposed
how the different chiral multiplets arise
from the brane construction.
The proposal
was examined in \cite{AH} by calculating superpotentials
for various brane configurations, and there was an agreement
with field theory expectation in all the cases.
In \cite{HZ} it was explained how chiral multiplets can be
localized in ten-dimensional space-time, by making use of the fact
that the theories in question actually live in five dimensions
with one direction being in a finite interval.
The chiral matter is localized on one boundary of the interval,
injecting an anomaly which flows through the interval to be absorbed by 
the chiral matter of opposite chirality
which is localized at the other boundary.
The four dimensional theory is then anomaly free. However, this is not
a chiral theory in the ordinary sense.
For this, we need to realize chiral multiplet in a more general
representation of the gauge group rather than
just the (anti-)fundamental or the adjoint.

One of the aims of this paper is to examine these ideas by studying
$N=2$ (i.e. $(2,2)$) supersymmetric theories in two dimensions
that can be considered as the dimensional reduction of chiral
theories in four dimensions.
Chiral matter in four dimensions
does not lead to gauge anomalies upon dimensional
reduction to two dimensions and hence 
theories which would be anomalous in four dimensions
are consistent theories in two dimensions.
We construct such gauge theories using branes of the Type IIA superstring
following the proposals of \cite{BH},
and compare with what we know about these theories.
We will find considerable agreement between them, providing further
support for the proposal of chiral symmetry realization.

There are many interesting features in $N=2$ theories in two
dimensions, and many exact results have been obtained.
Moreover, some of these theories are even completely integrable.
In section 2, 
the general background for $N=2$
theories in two dimensions is summarized.
One motivation of the study of such well-understood
systems using branes is to translate interesting field
theory phenomena to the language of branes.
In such a way more phenomena can be captured using the branes in cases
where the field theory tools are not as powerful as in two dimensions.
In particular, we obtain a hint for realizing non-fundamental
representation of the gauge group.

However the interplay between gauge theories and branes goes both ways.
Another aim of this paper is to use brane configurations
to deepen the understanding of $N=2$ supersymmetric theories
in two dimensions.
Using the brane construction we get
new interpretation for phenomena which are not clear from current
methods in field theory.
In some cases we obtain also some predictions,
qualitative and quantitative, which were not known before.

In section 3, we construct brane configurations in Type IIA
string theory. We examine the space of vacuum configurations
in this set-up, and compare with the space of
classical vacua of the field theory.

In section 4 we solve the proposed models of section 3 by taking the strong
Type IIA coupling, going over to $M$ theory
using the methods of \cite{WittenM5}.
We see that it correctly captures important quantum effects, such as
running of the Fayet-Iliopoulos coupling and the anomaly
of an R-symmetry group.
We also show that it correctly reproduces the number of quantum vacua
together with the discrete chiral symmetry breaking,
in the cases corresponding to the $\CP^{n-1}$
and Grassmannian sigma models.
In particular, we show that the relation of the
quantum cohomology ring of the
$\CP^{n-1}$ model is realized in the $M$ theory picture.

In section 5, we discuss
continuation past infinite coupling which is realized by an
interchange of two NS branes \cite{HW,Kutasov}
and interpret it as a transition between two gauge theories
which leads to level rank duality \cite{Schnitzer,tsu} of WZW models.
Other brane motions lead to new transitions in two dimensions which are
discussed in section 5.1.

One more important viewpoint emerges from the brane construction.
The brane picture demonstrates
a relation between $p$-dimensional theories with 4 supercharges and
$p+2$-dimensional theories with 8 supercharges ($p\le4$). 
The relation is that the parameter space of the $p$-dimensional theory is the
moduli space of vacua of the $p+2$-dimensional theory.
This viewpoint was first introduced and emphasized in \cite{AH}
for the $p=3$ case;
The Coulomb branch of the 5d theory is then the space of real 
masses of the 3d theory,
while the Higgs branch of the 5d theory is the space of
complex mass parameters for the 3d theory.
In section 5.2, we will touch this
correspondence for the $p=2$ case.

The simplest model in two dimensions with $N=2$ supersymmetry
and with chiral
matter is the $\CP^{n-1}$ model.
This model was studied intensively in the past
mainly because of various features which serve as a toy model for QCD. This
model is asymptotically free and has a theta angle with instantons.
All these features were attractive for modeling similar phenomena in QCD. 
 From the construction of the $\CP^{n-1}$ model in terms of branes this is not
surprising. Actually the brane construction provides an explanation of why
the $\CP^{n-1}$ model is a good toy model for QCD.
One way of interpretation of 
the brane system is as a D2 brane propagation on the
world volume of a configuration of branes which realizes some limit of $N=1$
supersymmetric Yang-Mills theory with gauge group $SU(n)$.
The world volume theory which is realized on the finite D2 brane is the
$\CP^{n-1}$ model. Thus the D2 brane probes some of the features of $N=1$
supersymmetric Yang-Mills theory.

In section 6 we discuss the realization of solitons
in the $\CP^{n-1}$ model as the topological excitations of a membrane
in which new boundary circles are created.
Each boundary circle is interpreted as an open string end point
which carries a quantum number of the fundamental representation
of the flavor group $SU(n)$.
Note that a string has only two ends and thus can realize
only up to second rank tensor representation,
while a membrane can have many boundaries
and so higher rank tensor representations can be realized.
Indeed,
it turns out that the fundamental
solitons interpolating adjacent vacua have one boundary and
are in the fundamental representation
of the flavor group $SU(n)$, but
solitons interpolating $\ell$-separated vacua
have $\ell$ boundaries and
are in the $\ell$-th anti-symmetric representation of $SU(n)$.
These properties in fact agrees with the field theory knowledge.
For example,
the fundamental solitons of the $\CP^{n-1}$ model
are known to be the elementary chiral multiplets
in the fundamental of $SU(n)$ which corresponds in the brane picture
to the elementary open Type IIA strings.
The mass spectrum of $\CP^{n-1}$ solitons
is also computed
and it also agrees with the field theory results.
The analysis is generalized to the case with deformation by mass
term and we determine the mass spectrum of the solitons,
which could not be achieved by field theory argument.

\section{$N=2$ Theories in Two Dimensions}

~~~~~In this section, we describe general properties of $N=2$
supersymmetric field theories in two dimensions.
We also describe a gauged linear sigma model realization of
supersymmetric non-linear sigma models
\cite{W79,DDL,W93}. In particular,
we discuss the soliton spectrum of the $\CP^{n-1}$ model
and its deformation by mass term.

\medskip

\subsection{General Background on $N=2$ SUSY Field Theories}

~~~~~$N=2$ supersymmetry in two dimensions can be obtained
by dimensional reduction from $N=1$ supersymmetry in four dimensions.

$N=1$ supersymmetry algebra in four dimensions
contains four supercharges which transform as
Majorana spinors under $d=4$ Lorentz group
(one left handed and one right handed spinors which are conjugate to
each other).
The SUSY algebra contains one $U(1)$ R-symmetry
under which the left handed supercharges have charge $-1$
and the right handed ones have charge $+1$.

By dimensional reduction to two dimensions (i.e.
eliminating the dependence of fields on two coordinates $x^{2,3}$),
the four-dimensional Lorentz group is broken to the
two-dimensional Lorentz group and an internal symmetry group
associated with the rotations in the $x^{2,3}$ directions
which we call $\UA$.
A left (right) handed
spinor in four dimensions becomes one Dirac spinor in two dimensions
--- one left and one right handed spinors with opposite
$\UA$ charge $\mp 1$ ($\pm 1$).
The four supercharges
are thus two Dirac spinors $Q_{L,R}$ and
$\bQ_{L,R}$
($L,R$ denotes the
two-dimensional chirality and absence/presence of {\it bar}
indicates the four-dimensional chirality)
which carry $\UA$ charge $-1,+1$ and $+1,-1$ respectively.
These are related to each other under conjugation by
$(Q_L)^{\dag}=\bQ_L$ and $(Q_R)^{\dag}=\bQ_R$.
They obey the commutation relation
\beqa
&&\{Q_L,\bQ_L\}\,\,=\,\,2(H+P),\label{comL}\\
&&\{Q_R,\bQ_R\}\,\,=\,\,2(H-P),\label{comR}\\
{\rm and} && Q_L^2=Q_R^2=\bQ_L^2=\bQ_R^2=0\label{nil}
\eeqa
where $H$ and $P$ are Hamiltonian and momentum operators.
In the absence of the central charge which we will describe shortly,
the other commutators vanish.

The $U(1)$ R-symmetry in four-dimensions can reside in two dimensions
as another internal symmetry --- which we call $\UV$ ---
under which
the supercharges $Q_{L,R}$ and $\bQ_{L,R}$ carry
charge $-1,-1$ and $+1,+1$ respectively.
Thus,
there are two $U(1)$ R-symmetry groups, $\UV$ and $\UA$.
The action on the supercharges is exhibited as
\beq
\begin{array}{cc}
Q_R\,\,&\,\,\bQ_L\\[0.4cm]
Q_L\,\,&\,\,\bQ_R
\end{array}
\eeq
where the upper (lower) row is assigned a $\UA$ charge $+1$ ($-1$)
while the right (left) column is assigned a
$\UV$ charge $+1$ ($-1$).
Of course these R-symmetries can be broken explicitly
by a tree level superpotential or, in the quantum theory, by an anomaly.
A basic example for such effects is provided at the end of this subsection.

\bigskip
\noindent
\subsection*{\it Representations}

~~~~~The two basic representations of the $N=1$ SUSY algebra in
four-dimensions, (anti-)chiral and vector multiplets,
go down to the corresponding representations of the two-dimensional
$N=2$ SUSY algebra.

A chiral multiplet consists of one complex scalar field $\phi$
and a Dirac fermion $\psi_{L,R}$. The action of the two R-symmetries
is exhibited (together with its conjugate anti-chiral multiplet
consisting of $\phi^{\dag},\overline{\psi}_{L,R}$)
as
\beq
\begin{array}{cc}
\psi_R\!\!&\\
&\!\phi\\
\psi_L\!\!&
\end{array}
\qquad\quad
\begin{array}{cc}
&\!\!\overline{\psi}_L\\
\phi^{\dag}&\\
&\!\!\overline{\psi}_R
\end{array}
\eeq
where the $\UA$ charge of the scalar component is zero,
while the $\UV$ charge is not specified since it can be shifted
by a constant.
The chiral multiplet is represented in the $N=2$ superspace as a
chiral superfield $\Phi$ obeying \footnote{
We follow the convention of \cite{WB,W93} in which
$$D_{\alpha}=\frac{\partial}{\partial \theta^{\alpha}}
-i\bar{\theta}^{\alpha}
\left(\frac{\partial}{\partial x^0}+\epsilon_{\alpha}
\frac{\partial}{\partial x^1}\right),~~~
\bD_{\alpha}=-\frac{\partial}{\partial \bar{\theta}^{\alpha}}
+i\theta^{\alpha}
\left(\frac{\partial}{\partial x^0}+\epsilon_{\alpha}
\frac{\partial}{\partial x^1}\right),$$
$$Q_{\alpha}=\frac{\partial}{\partial \theta^{\alpha}}
+i\bar{\theta}^{\alpha}
\left(\frac{\partial}{\partial x^0}+\epsilon_{\alpha}
\frac{\partial}{\partial x^1}\right),~~~
\bQ_{\alpha}=-\frac{\partial}{\partial \bar{\theta}^{\alpha}}
-i\theta^{\alpha}
\left(\frac{\partial}{\partial x^0}+\epsilon_{\alpha}
\frac{\partial}{\partial x^1}\right),$$
where $\alpha=L,R$ and $\epsilon_L=1,\epsilon_R=-1$.
Indices are lowered and raised by
$\epsilon_{\alpha\beta}$ with
$\epsilon_{LR}=1$ and its inverse $\epsilon^{\alpha\beta}$.
}
\beq
\bD_L\Phi=\bD_R\Phi=0,
\eeq
which can be expanded as
$
\Phi=\phi+\sqrt{2}\theta^{\alpha}\psi_{\alpha}
+\theta^{\alpha}\theta_{\alpha}F
$
where $F$ is a complex auxiliary field.

A vector multiplet consists of a vector field $A_{\mu}$,
Dirac fermions $\lambda_{L,R}$ and $\blambda_{L,R}$
which are conjugate to each other, and a complex scalar
$\sigma$ which comes from the $x^{2,3}$ components of the 
four-dimensional
vector field.
These are all in the adjoint representation of the gauge group.
The action of the two R-symmetry groups is
given by
\beq
\begin{array}{ccc}
&\!\sigma\!&\\
\blambda_L\!\!&&\!\!\lambda_R\\
&\!A_{\mu}\!&\\
\blambda_R\!\!&&\!\!\lambda_L\\
&\sigma^{\dag}&
\end{array}
\eeq
where the vector field $A_{\mu}$ is neutral under both.
The vector multiplet is represented in the $N=2$ superspace as
a vector superfield $V$ satisfying $V^{\dag}=V$
which can be expanded in the Wess-Zumino gauge as 
\beqa
V&=&\theta^L\bar{\theta}^L(A_0+A_1)+\theta^R\bar{\theta}^R(A_0-A_1)
-\theta^R\bar{\theta}^L\sigma-\theta^L\bar{\theta}^R
\sigma^{\dag}\nonumber\\
&&-i\theta^{\alpha}\theta_{\alpha}\bar{\theta}^{\beta}\blambda_{\beta
}
+i\bar{\theta}^{\alpha}\bar{\theta}_{\alpha}\theta^{\beta}
\lambda_{\beta}
-{1\over 2}\theta^{\alpha}\theta_{\alpha}\bar{\theta}^{\beta}
\bar{\theta}_{\beta} D
\eeqa
where $D$ is a real auxiliary field.
The super field strength is
defined by
$\Sigma=\{\overline{\cal D}_L,{\cal D}_R\}/2$
where ${\cal D}_{\alpha}=\e^{-V}D_{\alpha}\e^{V}$
and $\overline{\cal D}_{\alpha}=\e^V\bD_{\alpha}\e^{-V}$.
This is a twisted chiral superfield:
\beq
\overline{\cal D}_L\Sigma={\cal D}_R\Sigma=0.
\eeq
The lowest component of $\Sigma$ is the complex scalar
field $\sigma$.

\bigskip
\noindent
\subsection*{\it D-term, F-term and twisted F-term}

~~~~~There are three kinds of supersymmetric couplings.

One is the D-term which can be expressed as
\beq
\int \dd^4\theta\, K
\eeq
where $\int \dd^4\theta$ is the integration over all the Grassmannian
coordinates $\theta^L,\theta^R,\bar{\theta}^L,\bar{\theta}^R$
and $K$ is some real combination of superfields.
The D-term appears in the following as the gauge or
matter kinetic term. This term is invariant under both $\UV$ and 
$\UA$.

Second one is the F-term
\beq
\int \dd^2\theta\,\, W=\int \dd\theta^L\dd\theta^R\,\,
W|_{\bar{\theta}^{L,R}=0} 
\eeq
plus its Hermitian conjugate. Here $W$ is a holomorphic combination
of chiral superfields and is called superpotential as usual. 
The F-term is always invariant under $\UA$, but is invariant under
$\UV$ only when it is possible to assign $\UV$ charge to chiral
superfields so that
the superpotential carries charge $2$.
The latter condition is equivalent to saying that $W$
is quasi-homogeneous of degree 2 with respect to $\UV$.
Note that even if $W$ is not quasi-homogeneous,
a discrete subgroup of $\UV$ can be unbroken.

Third one is the twisted F-term
\beq
\int \dd^2\widetilde{\theta}\,\, \widetilde{W}=
\int \dd\theta^L\dd\bar{\theta}^R\,\,
\widetilde{W}|_{\theta^R=\bar{\theta}^{L}=0} 
\eeq
plus its Hermitian conjugate.
Here $\widetilde{W}$ is a holomorphic combination of 
twisted chiral superfields and is called
twisted superpotential. This preserves $\UV$ but breaks $\UA$
unless $\widetilde{W}$ is quasi-homogeneous of degree 2
with respect to $\UA$.
In a gauge system, the Fayet-Iliopoulos
D-term $-r\int \dd^4\theta\, \Tr V$ and the theta term
$i\theta\,\Tr F_A/2\pi$ can be described by a single
twisted F-term with
\beq
\widetilde{W}_{{\rm FI},\theta}={i\,\tau\over 4}\,\,\Tr 
\Sigma
\label{FI}
\eeq
where
$
\tau=ir+\theta/2\pi.
$
Since (\ref{FI}) is homogeneous of degree 2, this does not break
the R-symmetry $\UA$. However, in a gauge system
$\UA$ is often broken by an axial anomaly as in non-linear sigma model
based on non-Calabi-Yau manifolds.
Again, even if $\UA$ is broken by an anomaly, a discrete
subgroup can remain unbroken.

\bigskip
\noindent
\subsection*{\it Central Extension and BPS Bound}

~~~~~Consider a massive $N=2$ SUSY field theory with a discrete set of 
vacua.
If we put such a system on the flat Minkowski space $\R^{1,1}$,
there can be solitonic states
in which the boundary condition of fields at the left spatial 
infinity
$x^1=-\infty$ (specified by one vacuum) is different from the one at
the right infinity $x^1=+\infty$ (specified by another vacuum).
As is well known \cite{WO}, in such a theory
the $N=2$ SUSY algebra admits a central extension associated with
the topological charge of the soliton sectors.
Note that the central extension is impossible in a theory with
unbroken $\UV$ and $\UA$ R-symmetry, since the central term
should commute also with R-symmetry generators \cite{HLS}.

Let us consider a massive theory in which $\UV$ is broken by
a superpotential.
For example, $N=2$ Landau-Ginzburg (LG) models
with non quasi-homogeneous superpotentials.
In addition to
(\ref{comL}), (\ref{comR}) and (\ref{nil}), the algebra reads as
\beqa
\{Q_L,Q_R\}=2Z,\,\,&& \,\,\{\bQ_L,\bQ_R\}=2Z^*,
\label{c1A}\\
\{Q_L,\bQ_R\}=0,\,\,&&\,\, \{\bQ_L,Q_R\}=0,\\
{[} F_{\rm A},Q_L {]} =-Q_L,~~ {[} F_{\rm A},Q_R {]} =Q_R,\!\!\!\!&&
{[} F_{\rm A},\bQ_L {]} =\bQ_L,~~ {[} F_{\rm A},\bQ_R {]} =-\bQ_R
\label{c3A}
\eeqa
where $F_{\rm A}$ is the generator of $\UA$ R-symmetry.
One of the most important consequence of this algebra is that the
mass of the particle in a sector with central charge $Z$ is bounded
from below by \cite{WO}
\beq
M\geq |Z|\,.
\label{bps}
\eeq
This follows from the positive semi-definite-ness of the
anti-commutator of $(H-P)Q_L-Z\bQ_R$ and its hermitian conjugate
$(H-P)\bQ_L-Z^* Q_R$. This bound is saturated for states
on which the condition
$(H-P)Q_L=Z\bQ_R$
(called {\it BPS condition})
is satisfied.
For energy-momentum eigenstates satisfying the
BPS condition,
$Q_L$ and $\bQ_L$ are proportional to $\bQ_R$ and $Q_R$ respectively,
and thus the SUSY multiplet consists of two
states rather than four. This is called a {\it BPS multiplet}.

In a LG model with chiral superfields $X=(X^1,\ldots, X^d)$
and Lagrangian
\beq
S=\int \dd^2x\dd^4\theta \,\,K(X,X^{\dag})
+\left(\,\int \dd^2x\dd^2\theta \,\,W(X)\,+\,{\rm h.c.}\,\right)\,,
\eeq
with a non quasi-homogeneous superpotential $W(X)$,
the vacua are the critical points of the superpotential, $\partial 
W=0$.
For a solitonic state in such a system, the LG field $X(x^1)$
satisfies the boundary condition such that
$X(x^1=-\infty)$ is one critical point,
say $a$, and $X(x^1=+\infty)$ is another one, say $b$.
Then, the central charge $Z_{ab}$ in such a solitonic sector is 
\cite{WO,FMVW}
\beq
Z_{ab}=2\left(W(b)-W(a)\right)\,.
\label{cch}
\eeq
Indeed we can see the BPS bound $M\geq 2|W(b)-W(a)|$
from a classical argument \cite{CV}.
Let $g_{i\bar j}=\partial_i\partial_{\bar j}K$ be the K\"ahler 
metric.
Then the energy of a static configuration is
\beqa
E&=&\int \dd x^1\left\{\,\,
g_{i\bar j}\partial_1X^i\partial_1\bar X^{\bar j}
+g^{i\bar j}\frac{\partial W}{\partial X^i}
\frac{\partial \bar W}{\partial \bar X^{\bar j}}
\,\,\right\}\\
&=&\int \dd x^1 \left|\,\,
\partial_1X^i-\alpha g^{i\bar j}
\frac{\partial W}{\partial \bar X^{\bar j}}\,\,
\right|^2+2{\rm Re}\Bigl(\alpha^*(W(b)-W(a))\Bigr)\,,
\eeqa
for any phase $\alpha$, $|\alpha|=1$.
The second term of the RHS is maximum if we choose
$\alpha$ to be the phase of $W(b)-W(a)$, and 
thus, we obtained the bound $E\geq 2|W(b)-W(a)|$.
Note that this bound is independent of the K\"ahler metric.
In LG theories, the K\"ahler potential gets quantum corrections,
but the superpotential is not. Thus, this bound is exact quantum
mechanically. It is important to note that for
a BPS saturated configuration
$
\partial_1X^i=\alpha g^{i\bar j}
\partial_{\bar j}\bar W
$,
the trajectory along the spatial
direction $x^1$ of the superpotential $W$ is a straight line
\beq
\partial_1 W=\frac{W(b)-W(a)}{|W(b)-W(a)|}
g^{i\bar j}\partial_i W\partial_{\bar j} \bar W,
\eeq
connecting the two critical values $W(a)$, $W(b)$ \cite{CV}.

\medskip
For a theory in which $\UA$ is broken (say, by an anomaly), the same 
thing
can be said.
In addition to (\ref{comL})-(\ref{nil}), the $N=2$ SUSY algebra reads
\beqa
\{Q_L,Q_R\}=0,\,\,&& \,\,\{\bQ_L,\bQ_R\}=0,
\label{c1V}\\
\{Q_L,\bQ_R\}=2\widetilde{Z},\,\,&&\,\,
\{\bQ_L,Q_R\}=2\widetilde{Z}^*,\\
{[} F_{\rm V},Q_L {]} =-Q_L,~~ {[} F_{\rm V},Q_R {]} =-Q_R,\!\!\!\!&&
{[} F_{\rm V},\bQ_L {]} =\bQ_L,~~ {[} F_{\rm V},\bQ_R {]} =\bQ_R
\label{c3V}
\eeqa
where $F_{\rm V}$ is the generator of $\UV$. As in the previous case,
there is a BPS bound
\beq
M\geq |\widetilde{Z}|,
\eeq
with the BPS condition being $(H-P)Q_L=\widetilde{Z}Q_R$.
In a LG model for twisted chiral LG fields with a non quasi-homogeneous
twisted superpotential $\widetilde{W}$,
the central charge in a solitonic sector
is again given by the difference of the
critical values of $4\widetilde{W}$ at the two spatial infinities.
(The extra factor $2$ is due to the convention.)
In \cite{CV1,CV}, it has been argued that for any massive $N=2$ 
theory, one can define some kind of superpotential
(``holomorphic function'' on the discrete set of vacua)
such that the central charge in a solitonic sector is the
difference of the values of the superpotential.
As we will see however, in a theory with continuous Abelian symmetries
in addition to the R-symmetry, there can be a contribution to the
central charge from
charges of such Abelian groups:
\beq
Z\,=\,
2\Delta W +\sum_i m_i S_i~~~\mbox{or}~~~
\widetilde{Z}\,=\,4\Delta\tilW+\sum_{i}\tilm_iS_i
\label{Zmass}
\eeq
where $S_i$ are the Abelian charges
and $m_i$ or $\tilm_i$ are parameters such that
the Abelian symmetries are enhanced to some non-Abelian symmetry as
$m_i\to 0$ or $\tilm_i\to 0$ (typically mass parameters).
Note that the central charge
in such a case
is not determined just by the asymptotic condition at the spatial
infinities.

\bigskip
\noindent
\subsection*{\it Mirror Symmetry}

~~~~~As noted in \cite{LVW}, there is an interesting automorphism
of the $N=2$ supersymmetry algebra given by
\beq
\begin{array}{c}
F_{\rm A}\longleftrightarrow F_{\rm V}\\[0.2cm]
Q_R\longleftrightarrow \bQ_R
\end{array}
\label{mirror}
\eeq
with other generators kept intact.

Two $N=2$ theories are said to be mirror to each other
when there is an identification under which the $N=2$ SUSY generators
are mapped according to the above automorphism.
Such mirror pairs were first found and used effectively
in the study of $N=2$ super\,{\it conformal}
field theory associated with the sigma models with Calabi-Yau target
space (see \cite{Yau} and references therein). \footnote{
In fact, the automorphism (\ref{mirror}) extends to an automorphism
of the infinite $N=2$ superconformal algebra.}

Although it has not extensively been noted in the past,
a notion of mirror symmetry exists also for massive $N=2$
theories. In such a case, the automorphism involves also
the central elements. Look at the two types of massive theories
considered right above, where one type has unbroken $\UA$
and the other has unbroken $\UV$. Then, (\ref{mirror})
together with
\beq
Z\longleftrightarrow \widetilde{Z}
\eeq
defines an isomorphism
of the $N=2$ algebras with central extension:
(\ref{comL})-(\ref{nil}), (\ref{c1A})-(\ref{c3A})
is mapped to (\ref{comL})-(\ref{nil}), (\ref{c1V})-(\ref{c3V}).
The basic example of mirror symmetry in massive theories is
the pair of $N=2$ Sine-Gordon theory and supersymmetric
$\CP^1$ sigma model,
or more generally, the pair of $N=2$ $A_{n-1}$ affine Toda field theory
and SUSY $\CP^{n-1}$ sigma models.
In the affine Toda field theory $\UV$ is broken
by superpotential to $\Z_{2n}$
which is further broken spontaneously to $\Z_2$, while
in the $\CP^{n-1}$ model $\UA$ is anomalously broken to
$\Z_{2n}$ which is again further broken spontaneously to $\Z_2$.
Under the identification, the spontaneously broken discrete
$\Z_n$ symmetries are also mapped to each other.
Equivalence of soliton spectrum and S-matrices
were observed in \cite{FI} where we need to take a certain limit
of the coupling in the affine Toda side.
The mirror pair is recently generalized
to pairs of other kind of target space of positive first Chern class
and affine Toda-type field theory in the study of
twisted $N=2$ theories coupled to gravity \cite{EHX}
(see also \cite{Giv}).

\bigskip

\subsection{Gauged Linear Sigma Models}

~~~~~Let us consider $N=2$ supersymmetric
$U(k)$ gauge theory in two dimensions
with $n_1$ chiral multiplets $Q^{i}$
in the fundamental representation ${\bf k}$
and $n_2$ chiral multiplets $\tilQ_{\tilj}$
in the anti-fundamental representation ${\bf \overline{k}}$
($i=1,\ldots,n_1$; $\tilj=1,\ldots, n_2$).

The kinetic term of the Lagrangian of the theory is given by
\beq
{\cal L}_{kin}={1\over 4}\int\dd^4\theta
\left(\,
Q^{\dag}\,\e^{2V}Q+\tilQ\,\e^{-2V}\tilQ^{\dag}
-{1\over 2e^2}\Tr(\Sigma^{\dag}\Sigma)
\,\right)\,,
\eeq
where $\Sigma$ is the twisted chiral superfield representing the
field strength of the $U(k)$ vector superfield $V$
and
$e$ is the gauge coupling constant which has dimension
of mass.

In addition, we consider the Fayet-Iliopoulos (FI) and the theta terms
\beq
{\cal L}_{{\rm FI},\theta}
={i\,\tau\over 4}
\int\dd^2\widetilde{\theta}\,\,\Tr\Sigma
\,+\,h.c.
\eeq
where the FI parameter $r$ and the theta parameter $\theta$
are combined in the form
\beq
\tau=ir+\theta/2\pi.
\eeq

Also, we can consider the mass term
\beq
{\cal L}_{mass}=\sum_{i,\tilj}\int\dd^2\theta\,\,
m^{\tilj}_i\tilQ_{\tilj}Q^i\,+\,h.c.\,,
\eeq
where $\tilQ_{\tilj}Q^i$ is the natural gauge invariant
combination.
$m^{\tilj}_i$ are complex parameters which we call the
{\it complex masses}. Note that this term can be considered as
coming from the mass term which already exists in four dimensions.

Actually, there is another kind of mass term which
cannot be considered as coming from any coupling in four-dimensional
theories. This can be obtained by first gauging the flavor symmetry
$U(n_1)\times U(n_2)$ and giving a background value
to the scalar component of the vector superfield, and then
setting the fields to be vanishing.
This can be written as
\beq
{\cal L}_{\widetilde{mass}}
=\int\dd^4\theta\left(\,
Q^{\dag}\,\e^{2V_1}Q+\tilQ\,\e^{-2V_2}\tilQ^{\dag}
\,\right)
\eeq
where $V_1$ and $V_2$ are given by
\beq
V_1=\theta^R\bar\theta^L\tilm+h.c.,\qquad
V_2=\theta^R\bar\theta^L\btilm+h.c.\,.
\eeq
This preserves $N=2$ supersymmetry if and
only if $\tilm$ and $\btilm$
are (independently) diagonalizable:
\beq
\tilm=\left(
\begin{array}{ccc}
\tilm_1&&\\
&\ddots&\\
&&\tilm_{n_1}
\end{array}
\right)\,,
\qquad
\btilm=\left(
\begin{array}{ccc}
\btilm_1&&\\
&\ddots&\\
&&\btilm_{n_2}
\end{array}
\right)\,.
\eeq
We call these the {\it twisted masses}.
This is the two-dimensional version of the ``real mass term''
which were considered in \cite{DHOY,AHISS,DHO}.
Note that the shift of $\tilm$ and $\btilm$ by matrices
$c{\bf 1}_{n_1}$ and $c{\bf 1}_{n_2}$ proportional to
identity matrices can be absorbed by a redefinition
of the $\sigma$ field, and thus is irrelevant.

\bigskip
\noindent
\subsection*{\it The Space of Classical Vacua}

~~~~~After integrating out the auxiliary
fields,
the potential energy of this system is
\beqa
U&=&
{e^2\over 2}\,\Tr\left(\,QQ^{\dag}-\tilQ^{\dag}\tilQ-r\,\right)^2
\,+\,{1\over 8e^2}\,\Tr\,[\,\sigma,\sigma^{\dag}]^2
\nonumber\\[0.05cm]
&&\,\,+\,\,{1\over 2}\Bigl\Vert\,\sigma Q-Q\tilm\,\Bigr\Vert^2\,+\,
{1\over 2}\Bigl\Vert\,\sigma^{\dag}Q-Q\tilm^{\dag}\,\Bigr\Vert^2\,+\,
\Bigl\Vert\, Qm\,\Bigr\Vert^2
\nonumber\\
&&\,\,+\,\,{1\over 2}
\Bigl\Vert\,\tilQ\sigma-\btilm\tilQ\,\Bigr\Vert^2\,+\,
{1\over 2}
\Bigl\Vert\,\tilQ\sigma^{\dag}-\btilm^{\dag}\tilQ\,\Bigr\Vert^2\,+\,
\Bigl\Vert \,m\tilQ\,\Bigr\Vert^2\,.
\eeqa
We describe
the space of classical vacua
which is the space of zeros of $U$ modulo gauge transformations.
First of all, for the second term to be vanishing
$\sigma$ must be diagonalizable:
\beq
\sigma=\left(
\begin{array}{ccc}
\sigma_1&&\\
&\ddots&\\
&&\sigma_k
\end{array}
\right).
\eeq
The structure of the whole space
depends of the parameters $r$, $m$, $\tilm$ and $\btilm$.
\begin{description}
\item[(i)\,\,]
~When all these parameters are turned off,
the space of classical vacua is a singular space in which
there are roughly
two branches: In one branch
(``Coulomb branch''),
$\sigma$ is a non-zero
diagonal matrix and
$Q=\tilQ=0$,
while in the other branch
(``Higgs branch''),
$\sigma=0$ but $Q$ and $\tilQ$
can take non-zero values obeying $QQ^{\dag}=\tilQ^{\dag}\tilQ$.
Of course, there are ``mixed branches'' in which
first $l$ rows of $Q$ and first $l$ columns of $\tilQ$
are non-vanishing and
only the last $k-l$ of $\sigma_a$'s are non-vanishing.
Note that when $n_2=0$ (or $n_1=0$), the ``Higgs branch'' is trivial
$Q=0$ ($\tilQ=0$).
\item[(ii)\,\,]
When $r=0$ and a generic complex mass term is turned on,
the space of classical vacua consists only of ``Coulomb branch''
in which $Q=\tilQ=0$ and $\sigma$ is an unconstrained
diagonal matrix.
Upon specialization to $\rank m<{\rm min}\{n_1,n_2\}$,
``Higgs branch''
becomes possible.
\item[(iii)\,]
When $r=0$ and a generic twisted mass is turned on,
it is again only of ``Coulomb branch'', but when some of the
twisted masses for $Q$ and some for $\tilQ$ coincide,
there emanate ``Higgs branches'' at which some of $\sigma_a$'s are
tuned at the values of such twisted masses.
\item[(iv)\,]
When $r>0$ and all other parameters are turned off,
for the first term to be vanishing, $Q$ is non-zero and actually
must have rank $k$. This is possible only when $n_1\geq k$.
In this case, we must have $\sigma=0$
for the third term to be vanishing.
Thus, the space consists only of ``Higgs branch''.
Note that when $n_2=0$, this space is compact, that is, there
is no infinite ``flat direction''.
\item[(v)\,\,]
When $r<0$ and all other parameters are turned off, $\tilQ$
must have rank $k$
for the first term to be vanishing.
This is possible only when $n_2\geq k$.
In this case, $\sigma=0$
for the sixth term to be vanishing.
The space thus consists again only of ``Higgs branch''.
The same remark as (iv) applies to the case $n_1=0$.
\end{description}

\medskip
At the tree level,
fluctuation around each vacuum consists of massless
and massive modes which are tangent and transverse to
the space of classical vacua respectively.
The mass of the transverse modes depends on the
choice of vacuum and, in general, some massive modes become massless
at some special points such as the intersection of ``Higgs''
and ``Coulomb'' branches.
However, for the cases such as (iv) and (v),
the ${\rm mass}^2$ of the transverse modes are bounded from below
by a constant of order $e^2|r|$.
Thus, for this parameter region,
if we take the limit  $e^2\to\infty$
(or equivalently the long distance limit),
the massive modes
decouple and the system approaches
to a supersymmetric non-linear sigma model
whose target space is the corresponding space of
classical vacua. Namely the space of solutions of
\beq
QQ^{\dag}-\tilQ^{\dag}\tilQ=r
\label{KQ}
\eeq
modulo $U(k)$ gauge transformations.
In what follows, we mainly study such non-linear sigma models
realized by this gauged linear sigma models.

\bigskip
\noindent
\subsection*{\it Global Symmetry}

~~~~~The group of global symmetry of the system is at the tree level
$SU(n_1)\times SU(n_2)\times U(1)_a\times \UA\times \UV$,
where $\UA$ and $\UV$ are the two R-symmetry groups
and
$SU(n_1)\times SU(n_2)$ is the semi-simple part
of the flavor symmetry
group
$U(n_1)\times U(n_2)$ which acts on
$Q$ and $\tilQ$ as $k\times ({\bf n}_1,{\bf 1})$ and
$k\times ({\bf 1},\overline{\bf n}_2)$.
The vector combination of the center of
$U(n_1)\times U(n_2)$
is the same as the action of the center $U(1)$ of the gauge group
$U(k)$ and is not considered as a global symmetry.
The rest of the center is called $U(1)_a$ here.
If we turn on mass terms, part of these symmetries are explicitly
broken:
A generic twisted mass preserves the $U(1)$ symmetry
groups but breaks $SU(n_1)\times SU(n_2)$ to its
maximal torus.
A generic complex mass breaks $SU(n_1)\times SU(n_2)$ completely
but preserves $\UA$ and a combination of $\UV$ and $U(1)_a$.
These are restored by transforming the mass parameters in a suitable
way.
A choice of classical vacuum in the ``Higgs branch''
spontaneously breaks (part of) the flavor group
$SU(n_1)\times SU(n_2)\times U(1)_a$.

The above is a brief description of the symmetry
of the classical system. In the quantum theory, however,
there are two major corrections to what have been said.

One is the anomaly of $\UA$.
It acts oppositely
on the left and right handed fermions in each representation
of $U(k)$
and hence is generically anomalous.
Under the action of $\e^{i\alpha}\in \UA$, the fermion determinant
in a fixed gauge field $A$ changes by a phase shift
\beq
(2n_1-2n_2){i\alpha\over 2\pi}\int \Tr(F_A)
\label{shif}
\eeq
where $F_A$ is the curvature of $A$.
This shows that the $\UA$ R-symmetry is broken to its discrete
subgroup consisting of $2n_1-2n_2$ roots of unity:
\beq
\UA\longrightarrow \Z_{2(n_1-n_2)}\,.
\label{axlanm}
\eeq
In the case $n_1=n_2$, the whole $\UA$ is unbroken.
Note that in the general case $\UA$ can be restored by making it
act on the theta parameter by the shift $2(n_1-n_2)\alpha$
which absorbs (\ref{shif}).

In two-dimensional quantum field theory, a continuous symmetry cannot
be spontaneously broken. Therefore, even in the case when
some flavor symmetry
appears to be spontaneously broken at the tree level,
it must be restored in the full quantum theory unless it is broken by
an anomaly. This is the second correction to the classical
statement about the global symmetry.

\bigskip
\noindent
\subsection*{\it Renormalization}

~~~~~The theory is super-renormalizable with respect to the gauge
coupling constant $e$, as its mass dimension shows.
However, the FI parameter $r$ is dimensionless and this
must be renormalized due to a one loop ultra-violet divergence
which is present in the case $n_1\ne n_2$.

In order to see this, we look at the term in the Lagrangian
which depends linearly on the auxiliary field $D$ in the
vector superfield:
\beq
-\Tr\left\{D\left(QQ^{\dag}-\tilQ^{\dag}\tilQ-r\right)\right\}\,.
\eeq
The effective Lagrangian contains a term of this kind
in which $QQ^{\dag}-\tilQ^{\dag}\tilQ$ is replaced by its
expectation value. At the one loop level, the expectation value
is given by
\beq
\sum_{i=1}^{n_1}\int{\dd^2k\over (2\pi)^2}
{1\over k^2+\cdots}
-\sum_{\tilj=1}^{n_2}\int{\dd^2k\over (2\pi)^2}
{1\over k^2+\cdots}
\label{momint}
\eeq
where $+\cdots$ are terms depending on the complex/twisted masses
and background value of $\sigma$.
This integral is logarithmically divergent in the limit
$k\to\infty$ which is regularized by introducing a UV cut off
$\Lambda_{\rm UV}$.
There is no higher loop divergence, and the effective Lagrangian
can be made finite by renormalizing the bare FI parameter $r_0$
as
\beq
r_0=
{n_1-n_2\over 2\pi}\log\left({\Lambda_{\rm UV}\over\Lambda}
\right)\,,
\label{renoFI}
\eeq
to cancell the divergence of (\ref{momint}) as
$\Lambda_{\rm UV}\to \infty$.
Note that we are forced to introduced a dimensionful
constant $\Lambda$ which is an
analog of the dynamical scale of four-dimensional gauge theories.
The effective FI parameter at an energy scale $\mu$ is
then given by
$r(\mu)={n_1-n_2\over 2\pi}\log\left(\mu/\Lambda\right)$.
In other words, the theory for $n_1>n_2$ is asymptotically free
with respect to the coupling $g$ given by $r=1/g^2$, while
for $n_2>n_1$ it is so with respect to $g$ defined by $r=-1/g^2$.

This has an important implication.
Consider a theory with $n_1>n_2$.
It is always possible to find a scale $\mu$ at which $r(\mu)$
is positive: take $\mu$ to be much larger than $\Lambda$.
The space of
zeroes of the renormalized potential $U$ at that scale
is of the type (iv) in the above discussion,
and the theory is, in the limit $e^2\to\infty$,
interpreted as the non-linear sigma model with the target space
given by (\ref{KQ}).
The FI parameter $r=r(\mu)$ is interpreted as the size of the target
space, or equivalently its K\"ahler class.
Indeed, the way $r$ runs is exactly the same as the running
of the K\"ahler class of the sigma model, since
the one-loop beta function of the latter is given by the first Chern class
which, being equal to the anomaly of $\UA$,
is proportional to $n_1-n_2$.
Likewise, a theory with $n_2>n_1$ is always
of the type (v) in which
$r<0$ and is interpreted in the limit $e^2\to\infty$
also as the non-linear sigma model with the target space
given by (\ref{KQ}).
A theory with
$n_1=n_2$ is quite different and has separate ``phases''
corresponding to $r>0$ and $r<0$.
(See \cite{W93} for discussion of this type of theories.)

\bigskip
\subsubsection{The $\CP^{n-1}$ Model}

\medskip
~~~~~Here we consider the case
$k=1$, $n_1=n$, $n_2=0$ in some detail.
We first consider
the theory with $\tilm=0$
in which the global symmetry is $SU(n)\times \Z_{2n}\times \UV$.
The space of solutions of the D-term equation (\ref{KQ}) modulo
$U(1)$ gauge transformation
is the space of vectors in $\C^n$ of
${\rm length}^2=r$
modulo phase rotation, and hence is the same as
$n-1$ dimensional complex projective space
$\CP^{n-1}$.
Thus,
the theory describes
in the $e^2\to\infty$ limit the
supersymmetric sigma model with target space $\CP^{n-1}$
which has been studied extensively from various view points
\cite{W79,DDL,DDDS,KK,AL,CV2,FI,CV,EH,EHX}.

\bigskip
\noindent
\subsubsection*{\it Low Energy Effective Action}

~~~~~We consider integrating out the chiral superfield $Q$ and
obtain the effective Lagrangian as a functional of
the vector superfield. Due to the gauge invariance,
the result should be expressed in terms of
the twisted chiral superfield $\Sigma$ representing the
field strength.
The terms with at most two derivatives and not more than four fermions
can be written as
\beq
{1\over 4}\int\dd^4\theta\, K(\Sigma,\Sigma^{\dag})
+\left(\,\int\dd^2\widetilde{\theta}\,\tilW_{\it eff}(\Sigma)\,
+\,h.c.\,\right)\,.
\label{low}
\eeq
One can {\it exactly} determine
the effective superpotential $\tilW_{\it eff}(\Sigma)$ \cite{DDDS,CV}.
It is given by
\beq
\tilW_{\it eff}(\Sigma)
\,=\,{1\over 4}\left[\,i\,\tau\Sigma
-{n\over 2\pi}\Sigma\Bigl(\,\log\left(\Sigma/\mu\right)-1\,\Bigr)
\,\right]\,,
\label{tilWeff}
\eeq
where $\tau$ is the complex combination
$\tau=ir(\mu)+\theta/2\pi$
of the effective FI parameter
at the scale $\mu$ and the theta parameter.

The part of the effective Lagrangian (\ref{low})
which depends on the auxiliary field $D$ and the field strength
$F_A=F_{01}\dd x^0\wedge\dd x^1$ is
\beqa
&&
K_{\sigma \bar\sigma}\left(D^2+F_{01}^2\right)
+2\tilW_{\it eff}^{\prime}(\sigma)\left(D-iF_{01}\right)
+2\overline{\tilW_{\it eff}^{\prime}(\sigma)}\left(D+iF_{01}\right)
\nonumber\\
&&\,\,\, ={1\over 2\eeff^2}\left( D^2+F_{01}^2\right)-r_{\it eff}D
+{\theta_{\it eff}\over 2\pi}F_{01}
\label{low2}
\eeqa
where
$1/\eeff^2=2K_{\sigma\bar \sigma}$, and
$r_{\it eff}$ and $\theta_{\it eff}/2\pi$ are the
imaginary and the real parts of
\beqa
\tau_{\it eff}&=&i\,r_{\it eff}+{\theta_{\it eff}\over 2\pi}
\nonumber\\
&:=&-4i{\partial\tilW_{\it eff}(\sigma)\over\partial\sigma}
\,=\,\,\tau-{n\over 2\pi i}\log\left(\sigma/\mu\right)\,.
\label{eff2}
\eeqa
Integrating out the auxiliary field $D$, we get the
energy density
$\eeff^2r_{\it eff}^2/2$. Actually, there is a contribution
of the gauge field $F_{01}$
to the energy density. This follows from the fact \cite{Coleman}
that
the theta parameter $\theta_{\it eff}$ creates a constant 
electric field proportional to the minimum absolute value
$|\widetilde{\theta}_{\it eff}|/2\pi$
of $\theta_{\it eff}/2\pi+\Z$.
The contribution is then
$\eeff^2(\widetilde{\theta}_{\it eff}/2\pi)^2/2$.
Thus, the potential energy of the effective theory is
\beqa
U_{\it eff}&=&{\eeff^2\over 2}\left[\,
r_{\it eff}^2+\left(\widetilde{\theta}_{\it eff}\over
2\pi\right)^2\,\right]\\
&=&
{\eeff^2\over 2}\left|\,\,\widetilde{\tau}-
{n\over 2\pi i}\log\left(\sigma/\mu\right)\,\,\right|^2\,.
\eeqa
A supersymmetric vacuum is the zero of this potential
energy. There are $n$ such zeros.
One of them is given by
\beq
\sigma\,=\,\mu\exp\left({2\pi i \tau\over n}\right)
\,=\,\Lambda\e^{i\theta/ n}\,,
\eeq
and the others are obtained by the action of the discrete $\Z_{2n}$
subgroup of $\UA$ R-symmetry.
Namely, this discrete chiral symmetry is spontaneously broken to
$\Z_2$.

\bigskip
\noindent
\subsubsection*{\it Solitons in The $\CP^{n-1}$ Model}

~~~~~Since we have a discrete set of vacua, we expect
that there exist solitons which interpolate different vacua
at the two spatial infinities $x^1\to\pm\infty$.
If we only look at the effective Lagrangian
(\ref{low}), solitonic configurations are forbidden
by the Gauss law which is given by the
variation of (\ref{low}) or equivalently (\ref{low2})
with respect to $A_0$:
\beq
{\partial\over\partial x^1}
\left({1\over\eeff^2}F_{01}\right)
+{\partial\over \partial x^1}
\left({\theta_{\it eff}\over 2\pi}\right)=0\,.
\label{false}
\eeq
Recall from (\ref{eff2}) that
\beq
\theta_{\it eff}=\theta-n\arg(\sigma)\,.
\eeq
Then, integrating (\ref{false})
over the space coordinate $x^1$
and using the fact that $F_{01}=0$ for the vacua
at the spatial infinities $x^1\to\pm\infty$,
we see that
$
\arg\sigma(+\infty)-\arg\sigma(-\infty)=0
$,
which means that a configuration cannot interpolate
different vacua.

However, we can see that solitons {\it do} exist
if we take into account
the effect of the $n$ massive chiral multiplets
($Q^{i}$, $\psi^{i}$)
minimally coupled to the $U(1)$ gauge field $A_{\mu}$ as
$
-D_{\mu}Q^{\dag}D^{\mu}Q+i\bar\psi\gamma^{\mu}D_{\mu}\psi
$
\cite{W79}.
The Gauss law is then modified as
\beq
{\partial\over\partial x^1}
\left({1\over\eeff^2}F_{01}\right)
+{\partial\over \partial x^1}
\left({\theta_{\it eff}\over 2\pi}\right)+j^0=0\,,
\label{Gauss}
\eeq
where $j^0$ is the time component of the electric current
$j^{\mu}=iQ^{\dag}D^{\mu}Q-iD^{\mu}Q^{\dag}Q-\bar\psi\gamma^{\mu}\psi$
of the fields $Q^i$, $\psi^i$.
Then, integrating over the spatial coordinate, we have
\beq
\arg\sigma(+\infty)-\arg\sigma(-\infty)={2\pi\over n}\int \dd x^1 j^0\,.
\eeq

As a consequence of this identity, the {\it fundamental solitons}
which interpolate the neighboring vacua
$\sigma(-\infty)=\Lambda\e^{i\theta\over n}\to
\sigma(+\infty)=\Lambda\e^{{i\theta\over n}+{2\pi i\over n}}$
carry the electric charge $=+1$.
Indeed, these are the
elementary fields $Q^i,\psi^i$ which constitute BPS doublets
in the fundamental representation of the flavor group $SU(n)$
\cite{W79}.
Likewise, solitons interpolating vacua by
$\ell$ steps, $\sigma(-\infty)=\Lambda\e^{i\theta\over n}\to
\sigma(+\infty)=\Lambda\e^{{i\theta\over n}+{2\pi i\ell\over n}}$,
carry electric charge $\ell$. It is known that
the corresponding solitons consist of BPS saturated
bound states of $\ell$
elementary fields which transform as the $\ell$-th anti-symmetric
representation of $SU(n)$ \cite{KK,AL}.
The mass of such solitons is known to be \cite{KK,CV2,CV,EH}
\beq
M_{\ell}={n\over 2\pi}\Lambda |\e^{2\pi i\ell/n}-1|\,.
\label{solitonmass}
\eeq
This
coincides with what we naively expect
from the twisted
superpotential (\ref{tilWeff}):
Indeed, in spite of the presence of
the logarithm we can unambiguously determine the value of
$\tilW_{\it eff}$ 
at the $n$ vacua, by replacing $\tau$ in (\ref{tilWeff})
by $\widetilde{\tau}$. The value of $4\tilW_{\it eff}$
at the $j$-th vacuum
is $(n/2\pi)\mu\e^{2\pi i(\tau+j)\over n}$.
Then, (\ref{solitonmass}) is just the difference
$4|\Delta\tilW_{\it eff}|$
of the values at $j=\ell$ and $j=0$.

\bigskip
\noindent
\subsubsection*{\it Inclusion of Twisted Mass}

~~~~~Let us consider the theory with general twisted mass
in which the flavor symmetry $SU(n)$
is broken to $U(1)^{n-1}$.
Upon integrating out the chiral superfields $Q$,
we obtain the effective Lagrangian (\ref{low}) with the
twisted superpotential
\beq
\tilW_{\it eff}(\Sigma)
\,=\,{1\over 4}\left[\,i\,\tau\Sigma
-{1\over 2\pi}\sum_{i=1}^n
(\Sigma-\tilm_i)
\left(\,\log\Bigl({\Sigma-\tilm_i\over\mu}\Bigr)-1\,\right)
\,\right]\,.
\label{Weffm}
\eeq
Subsequent analysis is the same as in the case $\tilm=0$
and we have $n$ vacua corresponding to the $n$ roots
of
\beq
\prod_{i=1}^n(\sigma-\tilm_i)=\mu^n\e^{2\pi i t}.
\eeq
The soliton spectrum is also similar.
In the presence of a soliton with electric charge $\ell$,
the phase $\sum_{i=1}^n\arg(\sigma-m_i)$ changes by $2\pi \ell$.
As for the degeneracy,
there would be no much difference
from the case $\tilm=0$ at least for small $\tilm$,
and we expect ${n\choose \ell}$
solitons to exist in the $\ell$-th sector.

\begin{figure}[htb]
\begin{center}
\epsfxsize=2.5in\leavevmode\epsfbox{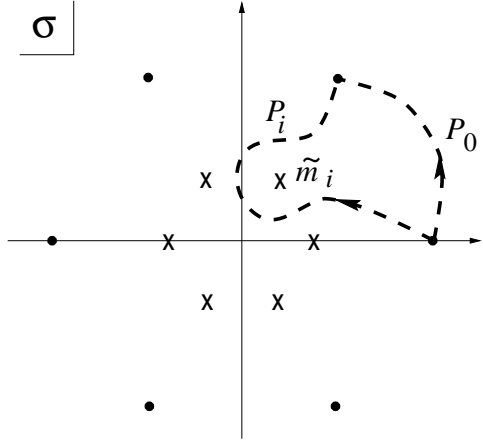}
\end{center}
\caption{The Two Paths}
\label{ambiguity}
\end{figure}

What are the masses of these solitons?
The theory with twisted mass has not been studied in the past.
A naive guess is the difference of the twisted superpotential
(\ref{Weffm}). However, in a theory with non-coincident
twisted mass,
we cannot
unambiguously determine the values of (\ref{Weffm})
at the vacua nor the difference
of the values at two vacua, as we now see.
Consider, for example, two nearby vacua
as depicted in Figure \ref{ambiguity}.
The difference of the values of $4\tilW_{\it eff}$
could be obtained by tracing the values of $\tilW_{\it eff}$ along
some path, say $P_0$, connecting them.
However, if we choose another path, say $P_i$, then
the difference changes by
\beq
\left.4\tilW_{\it eff}\right|_{P_i}
=\left.4\tilW_{\it eff}\right|_{P_0}-i\tilm_i\,.
\label{eqambi}
\eeq
So, there is an ambiguity in defining the superpotential,
and that is proportional to the twisted mass $\tilm$ and
vanishes in the $\CP^{n-1}$ model $\tilm=0$.
Actually, this ambiguity is related to the fact that there is a
continuous Abelian symmetry $U(1)^{n-1}$ for generic
values of $\tilm$.

We will see in section 6 using branes, that
the central charge is determined unambiguously
as the linear combination (as (\ref{Zmass})) of the
difference of the superpotential values
and the twisted masses times the $U(1)$ charges.
Indeed, the central charge is not determined just by the
boundary condition at the two spatial infinities.
The ambiguity of defining the superpotential cancels with
the ambiguity of the choice of $U(1)$ charges.
This is reminiscent of the formula for the central charge
in the $N=2$ theories in four dimensions \cite{SW2}
in which the ambiguity is related to the $SL(2,\Z)$ duality
of the low energy maxwell theory.
The central charge formula is thus modified, in the presence of
twisted mass terms by the amount
\beq
\sum_iS_i\tilm_i.
\eeq
A similar phenomenon appears also with real central charges in $N=2$
supersymmetry in three dimensions. See a discussion in \cite{AHISS}.

\bigskip
\noindent
\subsubsection*{\it
Relation to the $SU(n)$ $N=1$ super YM Theory in Four Dimensions}

~~~~~The supersymmetric $\CP^{n-1}$ sigma-model in two dimensions
has many properties in common with the $N=1$ supersymmetric
Yang-Mills theory in four dimensions with gauge group $SU(n)$.

Both have $n$ vacua with mass gap and have
discrete $\Z_{2n}$ chiral symmetry which is broken
spontaneously to $\Z_2$.
The $\CP^{n-1}$ model is asymptotically free with respect to
$g$ defined by $r=1/g^2$ as the $N=1$ Yang-Mills theory is
with respect to the gauge coupling constant $g_4$ where the one-loop
beta functions are both proportional to $n$.
The twisted superpotential of the $\CP^{n-1}$ model
is holomorphic with respect to the complex combination
$\tau=ir+\theta/2\pi$, while the superpotential of the $N=1$ Yang-Mills
theory is so with respect to the combination
$\tau_4=4\pi i/g_4^2+\theta_4/2\pi$
where $\theta_4$ is the theta parameter
in four dimensions.
Moreover, the effective twisted superpotential (\ref{tilWeff})
is the same
as the effective superpotential of the $N=1$ super YM theory
of \cite{VY} under the replacement $\Sigma\to S=W_{\alpha}^2$,
$\tau\to\tau_4$ and $\mu\to\mu^3$.

As a consequence, there is some close resemblance between the
solitons in the $\CP^{n-1}$ model and domain walls in the $N=1$
YM theory. In particular, both are BPS saturated.
Recently, there
is some interest in the study of domain walls in the $N=1$ YM theory
\cite{Shifman,SV} and the exact tension has been computed.
In particular, in \cite{W97} domain walls
in super YM theory are studied in the brane framework and
are claimed to be a D-brane for the QCD string. It would be
interesting to see the relation with the $\CP^{n-1}$ solitons
in the brane framework. The brane description of solitons in
the $\CP^{n-1}$
model is given in section 6.

There is another similarity of the two systems.
As noted in the previous subsection,
the $\CP^{n-1}$ model is dual under mirror symmetry to $N=2$
supersymmetric affine $A_{n-1}$ Toda field theory.
On the other hand, the $N=1$ $SU(n)$ super YM theory
is, when formulated on $\R^3\times S^1$, described
by a theory with chiral superfields with the superpotential being
the $A_{n-1}$ affine Toda potential
\cite{SW3d,KV,DHOY,AHISS}.

Relation of the two-dimensional systems
to four-dimensional gauge theories will be further discussed
in section 5 in terms of branes.

\bigskip
\subsubsection{Other $k=1$ Theories}

\medskip
~~~~~Let us consider a $U(1)$ gauge theory with general
$n_1,n_2$.
It is easy to obtain the effective twisted
superpotential for generic values of $\sigma$:
It is such that the effective FI and theta parameters are given by
\beqa
\tau_{\it eff}&=&-4i\,\,
{\partial \tilW_{\it eff}(\sigma)\over \partial\sigma}\nonumber\\
&=&\tau-{1\over 2\pi i}\sum_{i=1}^{n_1}
\log\left({\sigma-\tilm_i\over\mu}\right)
+{1\over 2\pi i}\sum_{\tilj=1}^{n_2}
\log\left({\sigma-\btilm_{\tilj}\over\mu}\right)
\label{effgen}
\eeqa
If the twisted masses are generic, there are ${\rm max}\{n_1,n_2\}$
vacua corresponding to the zeroes of (\ref{effgen}).
Something special
happens when some of $\tilm_i$ and some of $\btilm_{\tilj}$
coincide. Consider, for example,
the case with $\tilm_1=\btilm_1$. Then, there is a cancellation of two
terms in (\ref{effgen}). However, this does not mean that
the vacua are only the zeroes of (\ref{effgen}) (as many as
${\rm max}\{n_1-1,n_2-1\}$), because the fields
$Q^1$ and $\tilQ_1$ are massless at $\sigma=\tilm_1$
and the above description purely
in terms of $\sigma$ breaks down.
These massless fields span a one dimensional non-compact
complex manifold with a metric
$\dd z\dd \bar z/\sqrt{r+|z|^2}$
which is asymptotically $\C/\Z_2$,
where $r$ is the $r_{\it eff}$ at $\sigma=\tilm_1$.
Thus, the theory has ${\rm max}\{n_1-1,n_2-1\}$ vacua and the vacua of
the sigma model with such a target space.

Further discussion using branes on this is given in section 5.1.

\subsubsection{The Grassmannian Model}
\label{Grassmannian}

~~~~~Finally, we consider the case $k\geq 2, n_1=n,$ and $n_2=0$
with vanishing twisted mass $\tilm=0$.
The D-term equation (\ref{KQ}) for $Q=(Q^{ai})$
can be considered as requiring the $k$ vectors
$v_1,\ldots, v_k$ in $\C^n$ given by
$v_a=(Q^{ai})_{i=1,\ldots,n}$
to be orthogonal to each other
and have ${\rm length}^2=r$: $v_a^{\dag}v_b=r\delta_{a,b}$.
The $U(k)$ gauge transformation can be considered as the
unitary rotation of such orthogonal $k$-frames in $\C^n$
and does not change the $k$-plane in $\C^n$
which they span.
Namely, the space of classical vacua is the space
of $k$-planes in $\C^n$, which is the complex Grassmann manifold
$G(k,n)$.
Thus, the theory describes in the $e^2\to\infty$ limit
the supersymmetric non-linear sigma model with target space $G(k,n)$.

\bigskip
\noindent
\subsubsection*{\it Low Energy Effective Action}

~~~~~Like in the Abelian case, one can exactly perform
the integration over the chiral superfield $Q$.
Although a manifestly supersymmetric
form of the effective Lagrangian with respect to the
full field strength $\Sigma$
is not known, it is easy to determine the effective twisted
superpotential for
the case in which $\Sigma$ is diagonal,
$\Sigma={\rm diag}(\Sigma_1,\ldots,\Sigma_k)$.
It is given by
\beq
\tilW_{\it eff}(\Sigma)
\,=\,{1\over 4}\sum_{a=1}^k\left[\,i\,\tau\Sigma_a
-{n\over 2\pi}\Sigma_a\Bigl(\,\log\left(\Sigma_a/\mu\right)-1\,\Bigr)
\,\right]\,.
\label{WeffGrass}
\eeq
If the diagonal entries are well-separated
$|\sigma_a-\sigma_b|\,\,
\displaystyle{\mathop{>}_{\sim}}\,\,\Lambda$,
the off diagonal components of $\Sigma$ are heavy
and it is appropriate to integrate them out as well.
It does not give a contribution to $\tilW_{\it eff}$
since the off-diagonal components are in a complex and its conjugate
representations of the diagonal $U(1)$ gauge groups.
Thus, we may take (\ref{WeffGrass}) as the effective superpotential
in the region in which $\sigma_a$'s are well-separated.

Thus, a supersymmetric vacuum is at
$\sigma={\rm diag}(\sigma_1,\ldots,\sigma_k)$
where each entry $\sigma_a$ is one of the $n$-roots of
\beq
\sigma^n=\mu^n\e^{2\pi it}.
\label{sigman}
\eeq
Since the approximation is valid only when
$\sigma_a$'s are well-separated, distinct entries must be at
distinct roots of (\ref{sigman}).
The number of vacua is thus the number of possible selections of
$k$-elements among $n$-roots, namely,
\beq
{n\choose k}\,.
\label{nGrass}
\eeq

We have not considered the region in which some of the diagonal entries
are near one another, but it seems likely from the following reasons
that there is no supersymmetric vacuum in such a region.
First, if the distinct entries of $\sigma$ were allowed to coincide,
there would be supersymmetric vacua even in the case $k>n$
where there is no supersymmetric vacuum at the tree level.
Second, for the case $k\leq n$, the number (\ref{nGrass}) already
agree with the Euler number of the Grassmannian $G(k,n)$
which is the Witten index of the Grassmannian sigma model \cite{Windex}.
We will see in section 4 that this is consistent with
the observation in \cite{HW} that $s$-configurations
of the branes are not supersymmetric.

Given that these are the only supersymmetric vacua, we see that
the discrete chiral symmetry $\Z_{2n}$ is spontaneously
broken to some subgroups. A generic vacuum breaks it to $\Z_2$,
but some special vacua keep larger subgroups unbroken.

\bigskip
\noindent
\subsubsection*{\it Relation to $U(k)/U(k)$ Gauged WZW Models}

~~~~~It was shown in \cite{W93b} that the low energy limit, or the
dynamics of vacuum,
of the present model is described by the topological
field theory realized as the $U(k)$ Wess-Zumino-Witten (WZW) model
with the whole adjoint $U(k)$ group being gauged.
The level of the WZW model is $n-k$ for the $SU(k)$ part
and $n$ for the $U(1)$ part of the group $U(k)$. By Abelianization
of \cite{BT}, this gauged WZW model can be considered,
when formulated on the flat space-time,
as the
theory of $U(1)^k$ sigma model (represented by
$k$ free bosons $\phi_a$, $\phi_a\equiv\phi_a+2\pi$)
coupled to $U(1)^k$ gauge fields $A_a$ by
\beq
{n\over 2\pi}\sum_{a=1}^k\int \phi_a\dd A_a\,.
\eeq
The identification of the two systems
is essentially based on the observation that
this is the same
under $\phi_a={\rm arg}(\sigma_a)$
as the terms in
$\int\dd^2\widetilde{\theta}\,\,\tilW_{\it eff}+c.c.$
depending on ${\rm arg}(\sigma_a)$'s.

\bigskip
\section{A Type IIA Configuration}

~~~~~In this section, we construct brane configurations
such that the world-volume dynamics
at long distances
describes
$N=2$ supersymmetric gauge theories in two dimensions.
In particular, we propose configurations
which lead to the $N=2$ theories
obtained by dimensional reduction of
four-dimensional $N=1$ {\it chiral} gauge theories,
a typical example of which is the gauged linear sigma model
realizing the $\CP^{n-1}$ or Grassmannian sigma models.
In this and later sections, we provide evidence for the construction
and study such a gauge theory using the branes.

\medskip
Following \cite{HW}, the authors in \cite{Kutasov}
constructed a brane configuration in Type IIA string theory
whose world-volume dynamics at long distances
describes $N=1$ supersymmetric QCD in four dimensions. 
The configuration is in a flat space-time with time coordinate $x^0$
and space coordinates $x^1,\ldots,x^9$
and involves
two NS 5-branes spanning world-volumes in
the 012345 and 012389 directions,
$k$ parallel D4-branes stretched between them spanning world-volumes
in the 01236 directions, and $n$ parallel
D6 branes located between the two NS 5-branes
and spanning world-volumes in the 0123789 directions.
This yields $N=1$ supersymmetric $SU(k)$ QCD with $n$ flavors in four
dimensions.
One of the important features of such a theory
is the chiral flavor symmetry $SU(n)_L\times SU(n)_R$.
In a generic position of the D6 branes, we can only see the diagonal subgroup
$SU(n)$ in the
brane configuration, because
the configuration of D4 and D6 branes is locally of
theories with 8 supercharges. In
such a configuration, strings ending on the D4 and D6 branes
create the chiral multiplet $Q$ in the representation
$(\bk,{\bf n},{\bf 1})$
and the chiral multiplet
$\tilQ$ in the representation
$({\bf \overline{k}},{\bf 1},{\bf \overline{n}})$
{\it at the same point}, and therefore
it is also difficult to obtain chiral matter
in such a configuration.

\medskip
In the presence of the NS 5-brane spanning world-volume in the
012389 directions --- which we call the \NSp 5-brane ---
the configuration is locally of theories with 4 supercharges.
(In the absence of this brane the number of supercharges is 8.)
In particular, if the D6-branes and the \NSp 5-brane have the same
$x^6$ value, the D6-branes break into two pieces at the intersection
with the \NSp 5-brane \cite{BH}. The pieces in $x^7>x^7({\rm NS}^{\prime}5)$
will be called {\it upper-half} D6-branes, while the other pieces
will be called {\it lower-half} D6-branes.
In \cite{BH}, it is proposed that the strings ending on the D4 and 
the
upper-half D6-branes create the fundamental chiral multiplets $Q$
and the strings ending on the D4 and the lower-half D6-branes 
create the anti-fundamental chiral multiplets $\tilQ$.
Taking this for granted,
if we could take away the lower-half D6-branes from the
configuration, we would expect only the fundamental
chiral multiplet $Q$, leading to a chiral gauge theory.
However, in the absence of any other branes, such a configuration breaks
the charge conservation relation, and cannot be a stable
one, reflecting the gauge anomaly of the corresponding
four-dimensional
theory with only left-handed fermions in a complex
representation of the gauge group.
Indeed in \cite{HZ} it was argued how, in the presence of D8 branes,
a semi-infinite D6 brane can end on a NS fivebrane. This configuration
leads to a localization of a four dimensional chiral matter in ten
dimensional space time.
The presence of the chiral matter still induces an anomaly in the four
dimensional theory. However, the theory being really five dimensional
on an interval, gives rise to a flow of the anomalous current
along the five dimensional interval. This current is absorbed by a chiral
matter with an opposite chirality which is localized at the other end of
the interval.

\bigskip
Upon application of T-duality on the 23 directions,
we obtain a configuration of NS and \NSp 5-branes, $k$ D2-branes
with world-volume in the 016 directions,
and $n$ D4-branes with world-volume in the 01789 directions.
In this situation, when the D4-branes have the same $x^6$ value as the
\NSp 5-brane
and breaks at the \NSp into upper-half $x^7>x^7({\rm NS}^{\prime})$
and lower-half $x^7<x^7({\rm NS}^{\prime})$ pieces,
we can take away the lower-half pieces because D4-branes can
end on a NS 5-brane. This situation is different from the T-dual D6 - NS
configuration since now the D4 brane can move along the \NSp brane in
the 23 coordinates while the D6 brane can not move in any directions, its
boundary being the whole \NSp brane.
Moreover, two dimensional chiral multiplets do not give rise to anomalies
as their four dimensional analogs do. In this case the presence of a T-dual
to D8 brane, as in \cite{HZ} -- a D6 brane, is not necessary for localizing
the chiral matter in ten dimensional space time. This is reflected by the 
fact
that a D4 brane can end on a NS brane without having a D6 brane in the
background.
In this paper, we will study such a
configuration. Namely, the configuration involves

\noindent
--- A NS 5-brane with world-volume 012345 located
at a point in the 6789 directions,

\noindent
--- A \NSp 5-brane with world-volume 012389
located at a point in the 4567 directions,

\noindent
--- $k$ D2-branes with world-volumes 016
stretched between the NS and \NSp 5-branes,

\noindent
--- $n_1$ upper-half D4-branes with world-volumes
01789 ending on the \NSp 5-brane from above
in the $x^7$ direction, and

\noindent
--- $n_2$ lower-half D4-branes with world-volumes
01789 ending on the \NSp 5-brane from below in the $x^7$
direction. 

\medskip
\noindent
The configuration preserves 4 supercharges among the 32
of Type IIA string theory.
This can be seen from direct calculation of the broken supersymmetries.
Another way to see this is to note that this configuration is obtained from
the configuration of \cite{Kutasov} by T-duality (and removing
a part of the branes).
 
We note that this configuration is invariant under the rotations in
the $01$,
$23$, $45$ and $89$ planes. These will be inherited as
the Lorentz invariance and as global
symmetries $U(1)_{2,3}$, $U(1)_{4,5}$ and $U(1)_{8,9}$
in the field theory on the D2-brane world-volume which is now 
explained.

\bigskip
\subsection{Field Theory on the D2-brane World-volume}

\bigskip
~~~~~First note that the D2-branes are finite in the $x^6$ direction,
and any of the other branes has more than one
(semi-)infinite directions in addition to the 01.
Thus, as in \cite{HW},
we study the dynamics of the D2-branes and consider
the positions of the other branes to be fixed parameters.
Since the brane configuration is invariant under 4 supercharges
--- two left and two right ---
the world-volume dynamics at long distances
describes an $N=(2,2)$ (or simply $N=2$) supersymmetric
field theory in two dimensions.
By the rotational invariance of the configuration,
at least classically, the theory possesses the Lorentz invariance and
global symmetry groups $U(1)_{2,3}$, $U(1)_{4,5}$ and
$U(1)_{8,9}$.

\medskip
Light fields on the D2-brane world-volumes are given as follows.

An open string ending on the D2-branes
creates $N=8$ $U(k)$ vector multiplet,
but $6$ out of $8$ scalars are killed by the boundary condition
at the NS and \NSp ends, and only an $N=2$ $U(k)$ vector multiplet, 
or
equivalently a $U(k)$ twisted chiral multiplet $\Sigma$
remains. The value
$x^2+ix^3$ of the D2-branes correspond to the
eigenvalues of the scalar component $\sigma$
of the twisted chiral multiplet.
On dimensional grounds, these must be related by
$x^2+ix^3=\elst^2\sigma$, where $\elst$ is the string length.
These scalar components transform in the vector
representation (i.e. charge 2) under $U(1)_{2,3}$.

As in \cite{BH}, we propose that strings ending on the D2-branes
and the upper-half D4-branes create fundamental
chiral multiplets $Q^{i=1,\ldots,n_1}$,
while strings ending on the D2-branes and
the lower-half D4-branes create anti-fundamental chiral
multiplets $\tilQ_{j=1,\ldots,n_2}$.
In addition to the arguments given in \cite{HZ}, we will collect more
evidence for this proposal
in the following sections by observing that it has
the right consequences expected from the field theory analysis. 
The scalar components of these chiral multiplets are
singlets under $U(1)_{4,5}$ but transform in
the spinor representation (i.e. charge 1)
under $U(1)_{8,9}$.

\medskip
The positions of the NS and \NSp and D4-branes give parameters of the
theory. Note that the position of the NS brane is specified by
its $x^{6,7,8,9}$-value, but
we may put these to be zero $x^{6,7,8,9}({\rm NS})=0$
by a choice of the center of the coordinate.
Also, the position of the \NSp brane is specified by
its $x^{4,5,6,7}$-value but we may put
$x^{4,5}({\rm NS}^{\prime})=0$ by a choice of the origin
of the $45$ plane because
the NS brane has world-volume in these directions.
The remaining parameters are
\begin{itemize}
\item
The $x^6$-value of the \NSp brane determines
the bare gauge coupling
constant of the $U(k)$ gauge theory
\beq
{x^6({\rm NS}^{\prime})\elst\over g_{\it st}}=1/e^2\,,
\eeq
where $g_{\it st}$ is the Type IIA string coupling constant.
\item
The $x^7$-value of the \NSp brane is the Fayet-Iliopoulos 
parameter of the $U(1)$ factor of the $U(k)$ gauge group
\beq
{x^7({\rm NS}^{\prime})\over g_{\it st}\elst}=-r\,.
\label{classicalFI}
\eeq
\item
The $x^2+ix^3$ value of the upper-half or lower-half
D4-branes are the twisted masses for the chiral multiplets
$Q^i$ or $\tilQ_{\tilj}$:
\beq
\Bigl.\elst^{-2}(x^2+ix^3)\,\Bigr|_{{\rm D4}_i}=\tilm_i,
\qquad
\Bigl.\elst^{-2}(x^2+ix^3)\,\Bigr|_{{\rm D4}_{\tilj}}=\btilm_{\tilj}.
\eeq
\item
In the original configuration, all the D4 branes end
on the \NSp brane. Thus, their $x^4+ix^5$ value should be zero.
However, if one of the upper-half D4 branes and one of the
lower-half branes rejoin, they can be separated from the \NSp branes
and thus, in particular can have a non-zero $x^4+ix^5$ value.
If $i$-th upper and $j$-th lower D4-branes rejoin,
the $x^4+ix^5$ value corresponds to the complex mass $m_i^{\tilj}$
which enters into the tree level superpotential as
\beq
m_i^{\tilj}\tilQ_{\tilj}Q^i.
\eeq
Note, however, that we cannot have the general superpotential
$\sum_{i,\tilj}m_i^{\tilj}\tilQ_{\tilj}Q^i$ in this brane set-up.
This is because the brane configuration
chooses some diagonal embedding of the
matrix $m_i^{\tilj}$.
The most general mass matrix is given by the Higgs branch of
a four dimensional theory as explained below.
\end{itemize}

\begin{figure}[htb]
\begin{center}
\epsfxsize=4.5in\leavevmode\epsfbox{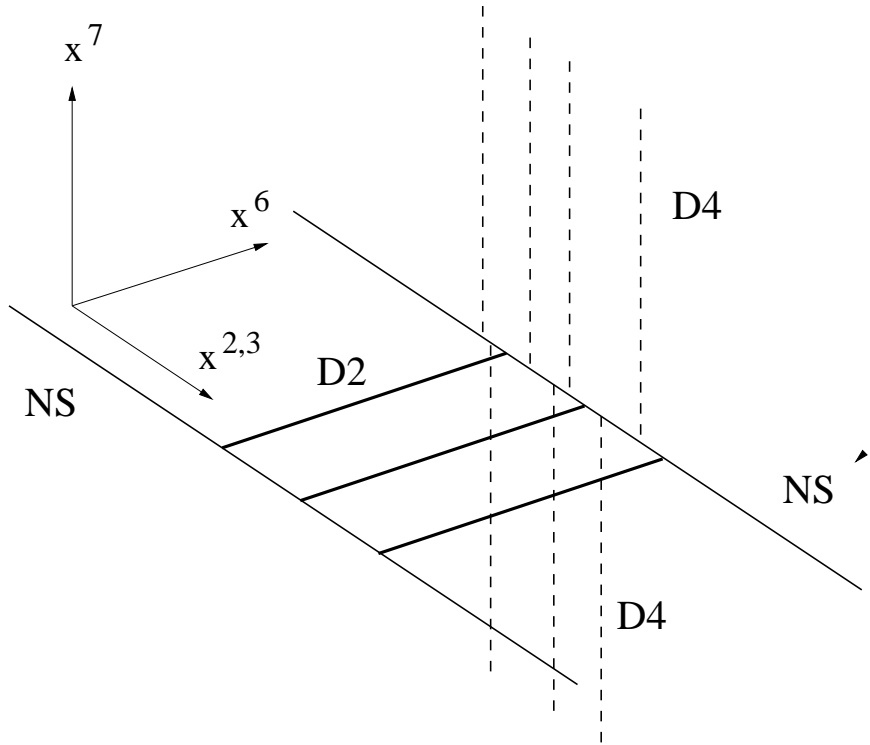}
\end{center}
\caption{The IIA configuration for general twisted mass}
\label{IIA}
\end{figure}

\medskip
In this set-up, it is easy to identify the chiral flavor
symmetry $U(n_1)\times U(n_2)$.\footnote{Note that
the axial and the vector $U(1)$ subgroups of this are the same as the
axial part of the group $U(1)_{4,5}\times U(1)_{8,9}$ and
the $U(1)$ subgroup of $U(k)$ gauge group respectively.}
The $n_1$ upper-half D4-branes are responsible for
the $U(n_1)$ factor and the $n_2$ lower-halves are
responsible for the $U(n_2)$ factor.
These are broken if the D4-branes are separated in the $23$
directions, that is, if the twisted masses $\tilm_i$ or $\btilm_j$
are turned on. In fact, the twisted masses can be interpreted as
the scalar component of the $N=2$ $d=4$ $U(n_1)\times U(n_2)$
vector multiplet of a gauge system
associated with the system of these D4 branes and the
\NSp 5-brane where,
the branes being semi-infinite in the $x^7$ direction,
the gauge dynamics is frozen.

Let us examine the situation from the point of view of a four dimensional
observer which lives in the common directions, 0189, of the D4 and NS$'$ 
branes.
 From the point of view of such an observer, the theory in question is a 
weakly
coupled $SU(n_1)\times SU(n_2)$ gauge theory with $N=2$ supersymmetry in four
dimensions. In addition there are hypermultiplets which transform in the
bi-fundamental representation. This consists of two chiral multiplets in the
$(n_1,\bar n_2)$ and $(\bar n_1, n_2)$ representations.
Such a theory is known to have a moduli space of vacua which
contains two branches. The Coulomb branch of this theory is $n_1+n_2-2$ 
complex
dimensional while in the Higgs branch, we can turn on a matrix of rank which 
is
at most ${\rm min}\{n_1,n_2\}$.
We want to identify the moduli of this theory with the parameters of the two
dimensional theory. This identification is possible due to the hierarchy in
scales that exists in the problem since the moduli of the four dimensional
theory are associated with motion of the heavy branes which are slowly 
varying
compared to the D2 branes.

The Coulomb branch of the four dimensional theory is identified with the 
motion
of the D4 branes along the 23 directions and thus gives the twisted mass
parameters $\tilm_i$ and $\btilm_j$.
The transition from the Coulomb branch to the Higgs branch of the four
dimensional theory is done by reconnecting D4 branes from both sides of the
NS$'$ brane and moving them in the 456 directions. Thus the Higgs branch
of the four dimensional theory parametrizes the complex mass parameters,
$m_i^{\tilj}$, of the two dimensional theory.
Note that since the Higgs branch contains four real scalars for a
hypermultiplet and the complex mass parameters are only two real, there are 
two
additional parameters in the four dimensional theory, per one hypermultiplet.
These parameters correspond to the $x^6$ position of the D4 branes and to the
reduction of the D4 gauge field in the $x^7$ direction, $A_7$.
There are some examples in which these parameters affect the low energy 
dynamics
of the two dimensional theory. See \cite{AH} for examples of such effects, 
the
discussion there being on a three dimensional analog of the system discussed
here.
It would be interesting to study further such effects.

\medskip
In summary, we list the fields and parameters of the theory,
together with the transformation properties under
the gauge group $U(k)$ and
the global symmetry groups.
Note that the symmetries
$U(1)_{2,3}$, $U(1)_{4,5}$, and $U(1)_{8,9}$.
can be considered as $U(1)_{\rm A}$, $U(1)_{\rm V}$, and another
$U(1)_{\rm V}$
R-symmetry groups respectively. The charges in the list
denote the charges of the scalar components of the superfields:
\beq
\begin{array}{ccccccc}
&U(k)&SU(n_1)&SU(n_2)&U(1)_{2,3}&U(1)_{4,5}&U(1)_{8,9}\\
\Sigma&{\bf adj}&{\bf 1}&{\bf 1}&2&0&0\\
Q_i&\bk&{\bf n_1}&{\bf 1}&0&0&1\\
\tilQ_j&{\bf \overline{k}}&{\bf 1}&{\bf \overline{n}_2}&0&0&1\\
r&{\bf 1}&{\bf 1}&{\bf 1}&0&0&0\\
\tilm_i&{\bf 1}&{\bf adj}\oplus{\bf 1}&{\bf 1}&2&0&0\\
\btilm_j&{\bf 1}&{\bf 1}&{\bf adj}\oplus{\bf 1}&2&0&0\\
m_i^{\tilj}&{\bf 1}&{\bf \overline{n}_1}&{\bf n_2}&0&2&0
\end{array}
\eeq

\medskip
\bigskip
\subsection{Classical Space of Vacua}
\label{Class}

~~~~~In this subsection we describe the brane realization
of the classical space
of vacua as described in section 2.2.
There are five cases discussed in that section depending on the values of the
parameters $r, m, \tilm$ and $\btilm$. In terms of branes these are the
positions in 7, 45 and 23, repectively, of the various branes
other than D2.
In the following, we analyze the vacuum configuration of the D2-branes
under various positions of other branes.
The numbering here is correlated with the one in section 2.2.

\begin{enumerate}
\item
When all parameters are turned off -- all the heavy branes
are at the
origin of the ten dimensional space (the NS$'$ brane is at
the origin of 45
and 7 and the semi-infinite D4 branes are at the origin of 23),
there are two
branches. The Coulomb branch
is parametrized by the 23 positions of the D2 branes between
the two NS branes.
The Higgs branch is parameterized by the positions of segments
of D2 branes which
break along the D4 branes and move in the 789 directions.
The two branches meet at the origin of moduli space.
There are mixed branches in which $l$ D2 branes break with
positions in 789
(Higgs) and $k-l$ D2 branes do not break and have arbitrary
positions in 23
(Coulomb). When there are no lower (upper) D4 branes
the D2 branes can not
break and there is no Higgs branch.
\item
When the NS$'$ is at $x^7=0$ and pairs
of lower and upper D4 branes form
into infinite D4 branes and move away
in the 45 directions there is a Coulomb
branch where the D2 branes are free to move
in between the two NS branes.
If only $m<\min\{n_1,n_2\}$ infinite branes
leave the 45 origin a Higgs branch
is possible.

There is a different branch which is not visible
in the field theory. This is the
motion of the infinite D4 branes in the $x^6$ direction.
Let $z_i$ denote the
$x^6$ positions of the infinite D4 branes and $t_1(t_2)$
denote the $x^6$
position of the NS (NS$'$) brane.
When $z_i<t_2$ for some $i$, the
analysis is the same and the parameter $z_i$ is irrelevant.
When $z_i<t_1$ there
is a D2 brane created between the D4 and
the NS brane \cite{HW} which still
makes the $z_i$ position irrelevant.
When $z_i>t_2$ there is a phase transition
and the states coming from the
D4 brane decouple from the system.
The number of massless multiplets is changed.
\item
When the NS$'$ is at $x^7=0$ and
the semi-infinite branes have arbitrary
23 positions there is again a Coulomb branch
given by arbitrary positions of
D2 branes in the 23 directions.
However when the 23 positions of a lower and an
upper D4 branes coincide they can combine and form
an infinite D4. Then,
a Higgs branch can emanate from that point
by breaking the D4 branes and moving
them along the 789 directions. The 23
values of the D4 branes must therefore coincide with
the 23 values of the D4
branes for such a breaking to happen.
\item
When the $x^7$ position of the NS$'$ brane
is negative (which corresponds to $r$ positive
due to the minus sign in (\ref{classicalFI})), 
the D2 branes are stretched between the NS brane and the upper-half
D4-branes. To avoid
s-configurations \cite{HW} the right ends of distinct D2 branes
are in distinct upper D4-branes.
For this we need $n_1\ge k$.
The D2 branes are fixed to the origin of 23
and hence there is only a Higgs branch.
When $n_2\ne 0$ the D2-branes can break at the infinite D4-branes
and the resulting segment can move in the 789 directions,
while when $n_2=0$ there is no direction of the D2-brane motion.
\item
For negative position of the NS$'$ brane the analysis is the same by
exchanging the lower and upper D4 branes.

\end{enumerate}
It is evident that the space of classical field theory vacua almost
agrees with the space of vacuum configurations of the D2-branes
in this Type IIA set-up.
A non-compact flat direction corresponds
to a direction of the D2-brane motion,
while a compact direction corresponds to the absence of such
direction.

\bigskip
\section{Quantum Properties Via $M$ Theory}

~~~~~The description in the previous section of Type IIA
brane configurations is missing basic properties of the
quantum theory
--- running of the FI parameter $r$,
the anomaly of the $U(1)_{\rm A}$
R-symmetry, and
the spontaneous breaking of its
non-anomalous discrete subgroup, which are present for theories with
$n_1\ne n_2$.
For instance,
the Type IIA configuration is
invariant under $U(1)_{2,3}$ which is identified
with $U(1)_{\rm A}$ and no sign of anomaly was observed.

In this section, we see that these properties are correctly 
reproduced
by considering the corresponding configuration in $M$ theory
where membranes and fivebranes are involved.
We will also see that the number of vacua agrees with the one
known in field theory for the case $n_2=0$ ($\CP^{n-1}$ or
Grassmannian model).

\bigskip
\subsection{$M$ Theory Description of the Configuration}

\medskip
~~~~~A system of branes ending on other branes cannot
be described by any conformal field theory of weakly
coupled superstrings.
In \cite{WittenM5} it was shown that the system of
D4-branes ending on NS 5-branes with two transverse directions
can be described as a smooth configuration of a single
fivebrane in $M$ theory on $\R^{10}\times S^1$
in the eleven-dimensional supergravity limit,
and that, for some purposes, it gives us nice or even exact
results of
the gauge theory on the brane in the long distance limit.
A D4-brane is a fivebrane wrapped on the $S^1$ and NS 5-brane is a
fivebrane at a point in the $S^1$. The configuration of a D4-brane
ending on a NS 5-brane is interpreted as a wrapped fivebrane merging
smoothly with an unwrapped fivebrane.
In the brane configuration given in the previous section,
the NS brane is described as a flat fivebrane in $M$ theory,
but the system of upper- and lower-half D4-branes ending on
the \NSp brane
is described by a curved fivebrane.
The D2-branes are described as membranes stretched between
the two fivebranes.
The dynamics of such open membranes ending on fivebranes
has not been fully understood.
However, as we will see in the next subsection,
there are features
of quantum theory
that can be captured correctly without the detailed
knowledge of the membrane dynamics,
e.g.,
renormalization of the FI parameter $r$ and anomalous
breaking of an R-symmetry.
In some cases,
the description by $M$ theory determines the vacuum structure
of the theory and, moreover,
enables us to obtain some exact information, such as soliton
spectrum and masses.
Conversely, it would be interesting to investigate the dynamics of
open membranes by making use of knowledge of
two-dimensional quantum field theory.

\medskip
We introduce the coordinate $x^{10}$ of the circle $S^1$
in the eleventh direction where $x^{10}$ is identified with
$x^{10}+2\pi$.
The flat eleven-dimensional space-time $\R^{10}\times S^1$
is given by the metric
\beq
\dd s^2=-(\dd x^0)^2+
\sum_{\mu=1}^9(\dd x^{\mu})^2+R^2(\dd x^{10})^2
\label{elevenmet}
\eeq
where $R$ is the radius of the circle $S^1$.
Recall that the string length $\elst$ and the eleven-dimensional
Planck length $\elel$ are related by
$\elel=g_{\it st}^{1\over 3}\elst$ where the string coupling constant
$g_{\it st}$ is given by $g_{\it st}^{2/3}=R/\elel$.

Below, we give an $M$ theory description
of the Type IIA configuration
in section 3 in some detail.

\bigskip
\noindent
\subsection*{\it The D4-Branes Ending on the NS\,${}^{\prime}$ Brane}

~~~~~In $M$ theory,
the system of upper and lower D4-branes ending on the \NSp brane
can be described as a single fivebrane
of the form $\R^4\times C$
where $\R^4$ spans the coordinate $x^{0,1,8,9}$ and
$C$ is a real two-dimensional surface 
embedded in the four-dimensional space spanning
the coordinates $x^{2,3,7,10}$
which is at a fixed position in the $x^{4,5,6}$ directions.
Since it preserves eight supercharges, it must be
holomorphically embedded with respect to some complex structure.
By introducing the complex coordinates
\footnote{We use for the complex combination of $x^{2,3}$
the same symbol $\sigma$ as the scalar component of
the twisted chiral superfield, as these are identified
in section 3.1. The factor
$\elst^{-2}$ is the tension of the fundamental string.}
\beq
\sigma=\elst^{-2}(x^2+ix^3),
\qquad
s=R^{-1}x^7+ix^{10},
\eeq
the $x^{2,3,7,10}$ part of the
space-time can be considered as a hyper-K\"ahler
surface with a flat K\"ahler metric
\beq
g=\elst^{\,4}|\dd\sigma|^2+R^2|\dd s|^2\,,
\eeq
and a holomorphic two-form
\beq
{\bf\Omega}\,=\,R\elst^{\,2}\,\,\dd\sigma\wedge\dd s
\,=\,\elel^{\,3}\,\,\dd\sigma\wedge\dd s
\,.
\label{Omst}
\eeq
The curve $C$ corresponding to
a straight \NSp brane is at a point in the $s$-cylinder and is
coordinatized by $\sigma$, while the curve for
a straight D4-brane is at a
point in the $\sigma$-plane and is coordinatized by $s$ modulo
a shift by $2\pi i$.
When an upper-half D4-brane ends on an \NSp brane at a point
$\sigma=\tilm$,
then, the D4-brane bends the \NSp brane in such a way that
$s$-value of the \NSp brane varies as a function of $\sigma$
as if the D4-brane plays the role of
a source of the Laplace equation for $s=s(\sigma)$:
$\partial_{\sigma}\partial_{\bar\sigma}s=-\delta(\sigma-\tilm)$.
Since the fivebrane becomes a D4-brane extending to
$x^7=+\infty$ as we approach
$\sigma=\tilm$, the real part of $s(\sigma)$ must diverge
at $\sigma=\tilm$ but it should not diverge nowhere else
because there is no other D4-branes.
Also, the imaginary part of $s(\sigma)$ admits a $2\pi$ shift
as we go around $\sigma=\tilm$, but
no other kind of multi-valued-ness is allowed.
A unique holomorphic function satisfying these properties
is
\beq
s=-\log(\sigma-\tilm)\,+\,{\rm constant}\,.
\eeq
Likewise, if we consider $n_1$ upper-half D4-branes at
$\sigma=\tilm_1,\ldots,\tilm_{n_1}$ and
$n_2$ lower-half D4-branes at
$\sigma=\btilm_1,\ldots,\btilm_{n_2}$,
the curve $C$ is described by
\beq
s=-\sum_{i=1}^{n_1}\,\log (\sigma-\tilm_i)
+\sum_{j=1}^{n_2}\,\log (\sigma-\btilm_j)+\,{\rm constant}\,.
\eeq
We recall that
we have chosen the coordinates so that the D4 and \NSp branes are at
$x^4=x^5=0$. Let $L$ be the $x^6$-value of them.
Then, by introducing a single valued coordinate of the
$s$-cylinder as $t=\exp(s)$,
the fivebrane is described by
\beqa
&&t\prod_{i=1}^{n_1}(\sigma-\tilm_i)
=q\prod_{\tilj=1}^{n_2}(\sigma-\btilm_{\tilj})\,,
\label{MNSp1}\\[0.2cm]
&&x^4=x^5=x^6-L=0\,,
\label{MNSp2}
\eeqa
where $q$ is a non-zero constant.

\bigskip
\noindent
\subsection*{\it The NS Brane}

~~~~~The NS brane is described in $M$ theory as just the 
flat fivebrane spanning the coordinates
$x^{0,1,2,3,4,5}$ which is located at a point in the remaining
directions.
In section 3, we have chosen the origin of the coordinates so that
it is located at $x^{6,7,8,9}=0$. Likewise, we can also
choose the origin so that the NS brane is at $x^{10}=0$.
Namely, the NS brane is the fivebrane at $s=0$,
or equivalently,
\beqa
&&t=1\qquad\mbox{and}
\label{MNS1}\\[0.2cm]
&&x^6=x^8=x^9=0.
\label{MNS2}
\eeqa

\bigskip
\noindent
\subsection*{\it The D2 Branes}

~~~~~The D2 branes are described as membranes stretched between
the two fivebranes. For a fixed time $x^0$,
it is basically an infinite strip where the two
boundaries are at the two fivebranes, namely, one is constrained by
(\ref{MNS1})-(\ref{MNS2}) and the other by
(\ref{MNSp1})-(\ref{MNSp2}).
There can be a ``topological excitation''
in which the membrane is not just a strip but can have additional
boundaries or genus. In such a case, we require any boundary to be
in one of the two fivebranes. Indeed, in section 6
we will see such excitations
as solitonic states of some models.

\bigskip
\subsection{Quantum Properties of the Theory}

\bigskip
\subsection*{\it The Fayet-Iliopoulos and Theta Parameters}

~~~~~We recall that in the Type IIA description
the difference $\Delta x^7=x^7({\rm NS}^{\prime})-x^7({\rm NS})$
of the $x^7$-values of the NS and \NSp branes
is interpreted as the FI parameter $r$ of the $U(1)$ part of the
gauge group. More precisely,
\beq
r=-{\Delta x^7\over g_{\it st}\elst}\,.
\label{IIAFI}
\eeq
However, in the present situation,
the \NSp brane does not have a definite $x^7$ value
and varies as a function of $\sigma$.
Correspondingly, $r$ varies as a function of $\sigma$
and we interpret this as the effective FI coupling
at the mass scale $|\sigma|$.
Since $g_{\it st}\elst=R$, it is given by
\beqa
r(\sigma)&=&-{\Delta x^7\over R}=-{\rm Re}(s)\\
&=&\sum_{i=1}^{n_1}\,\log |\sigma-\tilm_i|
-\sum_{j=1}^{n_2}\,\log |\sigma-\btilm_j|-\log |q|.
\eeqa
Indeed, at large $|\sigma|$ it behaves as
$r\sim (n_1-n_2)\log|\sigma|$, and this agrees with
what we expect from the renormalization (\ref{renoFI})
of the FI parameter,
up to a factor of $2\pi$ for which we have not been careful.
Moreover, for a suitable choice of $\log q$
this is exactly the same (modulo the $2\pi$ factor)
as the effective FI coupling $r_{\it eff}$ as a function of the
scalar component of the twisted chiral superfield $\Sigma$
(see (\ref{eff2}) and (\ref{effgen})).

In addition to this, we also interpret the separation
of NS and \NSp branes in the $x^{10}$ direction as
the theta parameter of the $U(1)$ part of the gauge group.
This can be understood in the following way.
Consider a (generically non-supersymmetric) configuration
of a membrane stretched between the fivebranes
where the boundary on the right has a fixed $x^{10}$ value
as well as the one on the left does.
Since the gauge field on the D2-brane
is dual to the scalar field
representing the position of the membrane in the eleventh
direction, we have $F_{01}=g_{\it st}\elst^{-1}\partial_6x^{10}$.
\footnote{
Actually, it is $F_{01}-B_{01}=g_{\it st}\elst^{-1}\partial_6X^{10}$
\cite{TownsendLec}.
In this situation, however, $B_{01}=0$
as the directions 01 being parallel to the fivebranes.
}
Namely,
\beq
F_{01}\,\sim\, g_{\it st}\elst^{-1}{\Delta x^{10}\over \Delta x^6}
\,\sim\, e^2\Delta x^{10}.
\eeq
On the other hand, we know from field theory \cite{Coleman}
that
the theta parameter $\theta$ creates a constant electric field
$F_{01}\sim e^2\theta$. Thus, we can identify the separation
$\Delta x^{10}$
as the theta parameter.

Altogether, the complex combination
$
\Delta s=\Delta x^7/R
+i\Delta x^{10}
$
as a function of $\sigma$ is interpreted as the
effective coupling constant
$2\pi i\tau_{\it eff}
=-2\pi r_{\it eff}+i\theta_{\it eff}$.
By (\ref{MNSp1}) and (\ref{MNS1}), it is given by
\beq
\Delta s=-\sum_{i=1}^{n_1}\,\log (\sigma-\tilm_i)
+\sum_{j=1}^{n_2}\,\log (\sigma-\btilm_j)+\log q.
\eeq
This agrees with the field theory knowledge
(\ref{eff2}),(\ref{effgen})
if we identify the constant $q$ as
\beq
q\,=\,\mu^{n_1-n_2} \e^{2\pi i \tau}
\,=\,\Lambda^{n_1-n_2}\e^{i\theta}\,.
\label{iden}
\eeq

\bigskip
\noindent
\subsection*{\it The Axial Anomaly}

~~~~~The Type IIA configuration has an invariance under
the rotations in the 23 plane.
Indeed, if we set all the twisted mass to be zero,
the configuration is invariant under the action of
$\e^{i\alpha}\in U(1)_{2,3}$,
$\sigma\to\sigma\e^{2i\alpha}$,
up to the positions of the D2-branes which depends on the choice of
vacua.
In the $M$ theory configuration, however,
the symmetry is reduced due to the modification
of the fivebrane on the right.
If we set $\tilm_i=\btilm_i=0$,
the fivebrane on the right is described by the equation
$t\sigma^{n_1}=q\sigma^{n_2}$, and this is invariant under
$\e^{i\alpha}\in U(1)_{2,3}$ only if $\e^{2i(n_1-n_2)\alpha}=1$.
Namely, $U(1)_{2,3}$ invariance is broken to its discrete
subgroup $\Z_{2(n_1-n_2)}$. This corresponds to the anomalous
breaking of the $\UA$ R-symmetry (\ref{axlanm}).
This discrete symmetry might be further broken by the configuration of
membranes, which corresponds to the spontaneous breaking
by a choice of vacuum.

\bigskip
\noindent
\subsection*{\it Validity of the Approximation}

~~~~~The theory on the branes
is in general different from the conventional
quantum field theory because the former
interacts with gravity and string excitations
in the bulk and involves
the modes associated with Kaluza-Klein reduction
on the interval.
If we are to draw some information on the two-dimensional
field theory from branes,
we must at least be able to find a limit in which all these extra
modes decouple from the system.

In the present context, there are essentially three parameters
that characterize the brane configurations
(for this part of the section,
we turn off the mass parameters): the separation
$\Delta x^6(=L)$ of the two fivebranes in the $x^6$ direction,
their separation $\Delta x^7$ in the $x^7$ direction
or the parameter $q$
that characterize the fivebrane on the right,
and the radius $R$ of the circle in the eleventh direction.
On the other hand, there are only two parameters that characterize
the field theory: the bare gauge coupling constant $e^2$ and the
FI parameter $r$ or the
scale parameter $\Lambda$ that organizes the running of $r$.
They are related by
\footnote{We thank correspondence with Andreas Karch on related issues.}
\beq
1/e^2\,=\,{|\Delta x^6|\elel^{\,3}\over R^2}\,,
\qquad
r\,=\,{|\Delta x^7|\over R}\,.
\label{combi}
\eeq
Since the theory is asymptotically free with respect to the coupling
constants $e$ and $1/r$, 
if $1/r$ is very small at the scales of gravity, string
and Kaluza-Klein excitations,
and $e$ is small compared to these scales,
we can neglect the effects of
these extra modes
at enough lower energies.
This condition is satisfied
if the radius $R$ is much smaller than $\elel$ and the other
parameters $|\Delta x^6|$, $|\Delta x^7|$.

Note that this is the weak coupling Type IIA limit of the $M$ theory
and is not a parameter region in which the low energy
supergravity approximation of $M$ theory is valid.
However, as far as
the qualitative features as well as
quantities that depends only on
the combinations (\ref{combi}) are concerned, it seems that
we can make some prediction and perform a computation by
going to a region in which
$R$
is large where we can use the eleven-dimensional
supergravity approximation.

There is, however, one thing which one must be careful.
The condition that $1/r$ is very small at the Planck scale
$\elel^{-1}$, the string scale $\elst^{-1}=R^{1/2}\elel^{-3/2}$
and the scale $1/|\Delta x^6|$ of Kaluza-Klein modes
is equivalent with
the condition that $\Lambda$ is much smaller than these energy
scales.
The identification (\ref{iden}) yields
$\Lambda=|q^{1/(n_1-n_2)}|$
and this is proportional to the characteristic length
$|\Delta x^{2,3}|=\elst^{\,2}|q^{1/(n_1-n_2)}|$
of the fivebrane on the right. In short,
\beq
\Lambda=R|\Delta x^{2,3}|/\elel^{\,3}.
\eeq
Thus, if we are to increase the radius $R$ beyond $\elel$
and keep $\Lambda$
to be small compared to the scales $\elel^{-1}$ and
$R^{1/2}\elel^{-3/2}$, then, the length
$|\Delta x^{2,3}|$
becomes very small (compared to $\elel$).
Since this length is the distance of the
different branches of the part of the
fivebrane on the right described by $\sigma^{n_1-n_2}t=q$,
if it is small, we are probing
the system of parallel and nearly coincident fivebranes.
Namely, the (2,0) superconformal field theory in six dimensions of
\cite{W20,Strominger}. Use of supergravity approximation
is therefore valid only if we increase $\Lambda$.
However, if the result of some computation
in this approximation
depends only on $\Lambda$ (and other quantities that appears
in the field theory) and is totally independent of
$\elel$, we can still expect the result to be
a good prediction. One such example in the past is the BPS mass
formula in the $N=2$ theories in four-dimensions \cite{WittenM5}
in which the distance of the fivebranes corresponding to the
D4-branes is of order $\Lambda$ and must be small
in the field theory limit, but should be large for the supergravity
approximation to be valid.
In the present paper, we compute
the mass of the BPS states of the $\CP^{n-1}$ model
and its deformation (see section 6).

\bigskip
\subsection{The Structure of Vacua}

\medskip
~~~~~A supersymmetric vacuum of the two-dimensional field theory
is realized as a configuration of membranes which preserves
four of the supercharges of $M$ theory.
The condition of four unbroken supercharges
is simply that each of the $k$ membranes are located at a
point in the directions transverse to 016.
Namely, the worldvolume of each membrane is a straight strip
$\R^2\times I$, where $\R^2$ is the 01 part of the space-time
and $I$ is a segment in $0\leq x^6\leq L$ located at a definite position
in the 2345789 and 10 directions. Since
it is stretched between the two fivebranes, 
the two ends of $I$ must satisfy the conditions
(\ref{MNS1})-(\ref{MNS2}) and (\ref{MNSp1})-(\ref{MNSp2}).

Below, we determine the vacua of the $\CP^{n-1}$
and Grassmannian models in this $M$ theory frame work.
These cases can be treated without detailed knowledge of
the dynamics of membranes, except that we need in the Grassmannian case
to make use of the rule \cite{HW} that $s$-configurations are not
supersymmetric.
We discuss other cases in the next section.

\bigskip
\noindent
\subsection*{\it The $\CP^{n-1}$ Model}

~~~~~We first consider configurations corresponding to
the $\CP^{n-1}$ model in which
$k=1$, $n_1=n$, $n_2=0$ and all the twisted masses are turned off.
In this case, the fivebrane on the right is described by (\ref{MNSp2})
and $\sigma^n t=q$.
Therefore, a configuration corresponding to a supersymmetric vacuum
is $\R^2\times I$ where $I$ is a segment in $0\leq x^6\leq L$
at $x^4=x^5=x^8=x^9=0$, $t=1$ and at one of the roots
of
\beq
\sigma^n\,=\,q\,.
\label{CPvac}
\eeq
Since (\ref{CPvac}) has $n$ roots, there are $n$ configurations
preserving four supersymmetries,
in agreement with the field theory result.
Also, it is evident that each choice breaks the discrete subgroup
$\Z_{2n}$ of $U(1)_{2,3}$ to $\Z_2$,
which corresponds to the spontaneous breaking
of the discrete R-symmetry group $\Z_{2n}\subset\UA$.

It is interesting to note that the relation
(\ref{CPvac}) is nothing but the quantum cohomology relation
of the $\CP^{n-1}$ model \cite{W89},
which represents the instanton correction
to the chiral ring.

\medskip
It is easy to extend this result
to the case where the twisted masses are turned on.
In this case, the equation (\ref{CPvac}) is modified as
\beq
\prod_{i=1}^n(\sigma-\tilm_i)=q\,.
\label{CPmvac}
\eeq
The number of vacua is still $n$
since the number of roots is, in agreement with 
the field theory result. It would be interesting to determine the
chiral ring of the model with twisted mass in the field theory frame
work. A natural guess is that it is described by (\ref{CPmvac}).

\bigskip
\noindent
\subsection*{\it The Grassmannian Model}

~~~~~The Grassmannian sigma model with target space $G(k,n)$
is described by $k$ membranes stretched between the fivebranes
in which $n_1=n$, $n_2=0$, and all $\tilm_i=0$.
A configuration preserving four supersymmetries
is a set of $k$ membranes $\R^2\times I_{(a)}$,
$a=1,\ldots, k$, where each $I_{(a)}$ is a segment in
$0\leq x^6\leq L$
at $x^4=x^5=x^8=x^9=0$, $t=1$ and at one of the $n$ roots
of (\ref{CPvac}).

The structure of vacua depends on whether two or more membranes
can be on top of each other. To answer this from first principles,
we need a detailed knowledge of the dynamics of open membranes
stretched between the fivebranes, which we do not have presently.
However, the configuration of coincident membranes in this set-up
is locally the same as the (T-dual of) $s$-configuration of \cite{HW}.
\footnote{To remind the reader,
$s$-configurations are configurations in which
$k$ D3 branes are stretched between a
NS fivebrane and a D5 brane in type IIB.
The statement is that for $k>1$ the configuration
breaks supersymmetry even
though apparently it need not.}
Indeed, since the fivebrane on the right is obtained from
the D4-branes ending on \NSp brane, we may view a point on it
as a point of one of the D4-branes.
Then, the configuration of coincident membranes can be viewed
as a configuration of two (or more) D2-branes stretched between
a single NS 5-brane and a single D4-brane, a T-dual of the
$s$-configuration of \cite{HW}.
Then, provided the rule
that $s$-configurations are not supersymmetric is correct, 
a configuration of two or more membranes on top of each other
is not supersymmetric.
It is plausible to admit this since the rule has passed through
various different checks \cite{HW,OV,HOO}.
It would be interesting to prove it directly
from the study of membrane
dynamics. 
We leave it as an open problem.

Thus, a configuration corresponding to a supersymmetric vacuum
is given by a choice of $k$ {\it distinct} roots among the
$n$ roots of (\ref{CPvac}). Thus, the total number of such
configuration is ${n\choose k}$, in agreement with
the field theory result. A choice of such configuration breaks
the discrete symmetry group $\Z_{2n}$.
The pattern of breaking $\Z_{2n}$
also agrees with what we expect from the
field theory.

It is interesting here also to note
that the quantum cohomology ring of the
Grassmannian $G(k,n)$ is the ring of symmetric
polynomials of $\sigma_1,\ldots,\sigma_k$ with each $\sigma_i$ obeying
the constraint (\ref{CPvac}) (see \cite{Ken,CV,W93b}).

We will discuss more about the Grassmannian model
in the next section.

\bigskip
\section{Continuation Past Infinite Coupling}

~~~~~Let us now turn to study non-trivial dynamics of the two dimensional 
theories at
hand by moving branes in space-time.
There is a very simple trick which can be applied to any brane 
configuration.
We can reorder positions of branes in the $x^6$ direction.
Let us review first what are the consequences of this operation in 
various
dimensions and supersymmetries.

In \cite{HW} a configuration of $N_c$ threebranes stretched between 
two
NS fivebranes together with $N_f$ D fivebranes was used to construct 
$N=4$
supersymmetric $U(N_c)$ gauge theory with $N_f$ flavors
in three dimensions.
It was shown there that
when the two NS fivebranes exchange their position the theory changes 
its matter
content. There is a phase transition to a $U(N_f-N_c)$ gauge theory 
with $N_f$
flavors. The gauge coupling of a given theory
is proportional to the inverse 
distance
between the two NS branes. Thus, the transition goes through infinite 
coupling for both gauge theories. In this sense we call one theory 
continuation
past infinite coupling of the other.
The transition is performed in the Higgs branch
of both theories and thus 
allows
us to study the equivalence between the Higgs
branches of the two theories.

The authors in \cite{Kutasov} applied
the same trick as in \cite {HW} 
to study
the dynamics of $N=1$ supersymmetric gauge theories in four 
dimensions.
For this case the exchange of the two NS branes leads to Seiberg 
Duality
\cite{Nati}. The two gauge theories which are involved 
are similar
to the theories in the three dimensional analog and are $SU(N_c)$ 
with $N_f$
flavors while the dual is $SU(N_f-N_c)$
with $N_f$ flavors and some 
mesons with a superpotential.
See a detailed discussion in \cite{Adam}.
Other applications of this effect can be found in
\cite{jon,BH,ted,tatar,AOT}.

Let us try to apply the trick of \cite{HW} to the system at hand.
In this case the number of matter fields is roughly half the number 
of fields
that we have in the two previous examples, however since the 
technique is the
same, we will get qualitatively equivalent results for this theory as 
well.
There are two ways to look at the problem. We have the type IIA 
picture which
describes the semiclassical limit of the theory and the $M$ theory 
picture which
describes the quantum behavior of the system. We will first describe 
the exchange of the NS branes in the type IIA picture.
Then we will describe the same transition in the $M$ theory setup and 
see that the
two pictures are really different. This serves as a good example for 
the quantum
correction of the transition and shows that it is a truly quantum 
effect.

\bigskip
\noindent
\subsection*{\it Type IIA Description of the Transition}

\begin{figure}[htb]
\begin{center}
\epsfxsize=2.5in\leavevmode\epsfbox{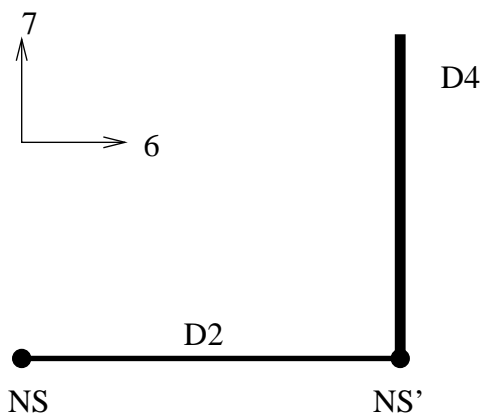}
\end{center}
\caption{$U(k)$ gauge theory coupled to $n$ chiral multiplets.
There are $k$ D2 branes stretched between NS and NS$'$ fivebranes.
$n$ semi-infinite (upper) D4 branes end on the NS$'$ fivebrane.
}
\label{cpn}
\end{figure}

~~~~~Consider the system
of branes as in section 3 with $n_1=n, n_2=0$ and 
$k<n$. We sketch this configuration in figure \ref{cpn}.
This configuration of branes describes $U(k)$ gauge theory with $n$ 
chiral
multiplets in the fundamental representation. As reviewed in section
\ref{Grassmannian}, in the infrared this theory was
shown to describe a sigma model on the space of Grassmannians $G(k,n)$.
\footnote{See, for example, \cite{Ken,W93b} and references therein.}

\begin{figure}[htb]
\begin{center}
\epsfxsize=2.5in\leavevmode\epsfbox{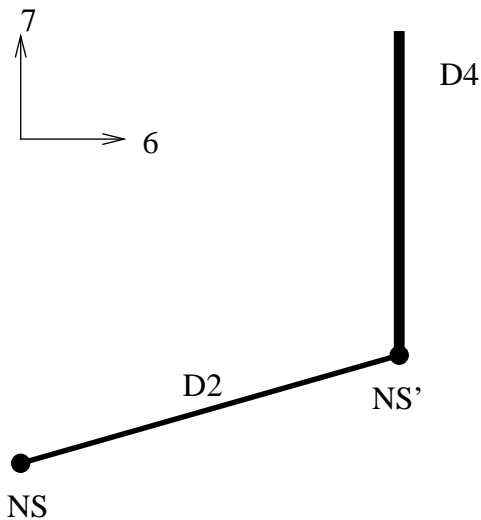}
\end{center}
\caption{Turning on a FI term when the D2 branes are in between
the two NS branes. The D2 branes must end on the NS branes to avoid
charge violation. As a result their orientation in space-time is
changed and thus supersymmetry is broken.}
\label{break}
\end{figure}

We want to exchange the $x^6$ position of the NS fivebrane and the
NS$'$ fivebrane. We need to avoid possible charge flow between the 
two
fivebranes when they pass. Therefore we want the fivebranes to avoid 
each other
in space time. For this we need to change the relative distance in 
the $x^7$
direction. This corresponds to turning on a FI parameter for the 
$U(1)$ part in
the $U(k)$ gauge group. However in a generic situation this has the 
effect of
breaking supersymmetry. Such a breaking is visible in the brane 
construction by
changing the orientation of the D2 branes in space. Indeed if we move 
the
NS fivebrane in the $x^7$ direction, when some D2 branes are 
stretched in
between the two fivebranes, the orientation of the D2 branes in space 
gets some
angle in the 67 plane and thus breaks supersymmetry.
This is sketched in figure \ref{break}.
We can avoid 
such a
breaking by letting the D2 branes end on other branes. The simplest 
way to do it
is to introduce D4 branes on which the D2 branes can end.
The only D4 branes present are the semi infinite branes which end on 
the NS$'$
brane. As such they can not move independently in the $x^6$ direction 
because
the NS$'$ brane is point like in this direction. An attempt to move 
the D4
branes away from the NS$'$ brane will result in RR charge violation.
However we can use lower
semi-infinite branes which come from infinity in the 23 directions and 
reconnect
with the upper D4 branes to form infinite D4 branes. Once they are 
infinite they
can leave the NS$'$ brane as there is no charge violation in this 
case. So we move $m$ lower D4 branes
from infinity to reconnect with $m$
upper D4 branes to form $m$ infinite D4 branes.
This can be done only if
$n\ge m$. Such a process corresponds to
turning on $m$ chiral multiplets
in the anti fundamental representation.
The 23 position of the lower
D4 branes being the twisted mass $\hat m_j$.
When the lower and upper branes reconnect
and the resulting infinite D4 brane
moves in the $x^6$ direction,
The chiral symmetry $SU(n)\times SU(m)$ is
broken explicitly.

\begin{figure}[htb]
\begin{center}
\epsfxsize=2.5in\leavevmode\epsfbox{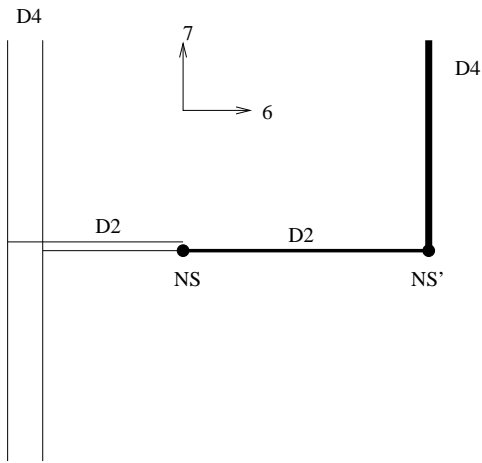}
\end{center}
\caption{A Type IIA description of the exchange of two NS branes.
This is a classical transition which is corrected quantum mechanically.
In the figure there are D2 branes stretched between infinite D4 branes
and a NS brane. The D2 branes are created when the D4 and the NS branes
pass each other in the $x^6$ direction.}
\label{infin}
\end{figure}

Next the D4 branes should cross the NS brane in the $x^6$ direction. 
They share
only one transverse direction and therefore can not avoid each other 
in space.
When they cross a D2 brane is created \cite{HW} which is stretched in 
between
them. If $m$ D4 branes cross the NS brane there are $m$ D2 branes 
which are
stretched in between the NS brane and each of the D4 branes. These D2 
branes are
not free to move. They are stuck on one side to the NS brane and on
the other side to the D4 branes.
We can still reconnect such branes with
the $k$ D2 branes that are
stretched between the two NS branes.
This is done by moving the latter branes
to touch the former branes which
are stack. Such a process corresponds
to changing the expectation values
of the adjoint matrix $\sigma$
in such a way that we go to a point where
the quark fields become massless.
Such a process can not be possibly correct
quantum mechanically because in two
dimensions there is no moduli space of
vacua, at most a discrete set of
vacua are allowed. Nevertheless let us
proceed with the semiclassical description.
In such a way
we reduce the number of branes attached
to the NS brane. If we do this for
all $k$ D2 branes between the two NS branes
there are no branes attached
to the NS brane left. At this stage
we can move the NS brane in the $x^7$
direction without breaking supersymmetry.
Note that the number of D4 branes, $m$,
must be greater or equal to $k$
for supersymmetry not to be broken.
 From the above considerations we see
that the minimal number of lower branes
needed for these processes to preserve
supersymmetry is $m=k\le n$.
So we choose this value and proceed.
The resulting configuration is
sketched in figure \ref{infin}.

At this point we can move the NS brane past
the NS$'$ brane by first moving
in the $x^7$ direction and then in the $x^6$ direction.
We encounter an apparent puzzle.
There are two ways to turn on the FI parameter, $r$.
One is to turn $r>0$ and
the other is to turn $r<0$. In the first case
the NS brane will meet the 
upper
D4 branes and when crossing will create a D2 for
each D4 brane. In contrast,
in the second case, the NS brane will not meet any
D4 branes and thus there will
be no creation of D2 branes at all.
This is in contradiction with the
expectation that the transition
will be independent of the path chosen!

We clearly see that the type IIA picture
has some ambiguity. We will see how
this is cured in the $M$ theory description of
this transition. For now let us
recall, as reviewed in section 2.2, that the FI
parameter gets renormalized in the quantum theory
(which is equivalent to going to $M$ theory limit).
Indeed the $r>0$ region which seems to be special
at the classical theory is
smoothed and can be continued to all values of $r$.
In terms of branes what happens is that
the D4 branes bend the five brane in
such a way that the $r>0$ region is extended.
So let us assume that the region
of $r>0$ is valid for every value of $r$.

With this assumption, when the NS brane crosses
the $n-k$ D4 branes there are
$n-k$ D2 branes created. The NS brane can then
go back to the origin of the $x^7$
direction and now the resulting theory
is $U(n-k)$ with $n$ fundamental 
fields.
Classically there seems to be some additional
matter and couplings however
the picture modifies in the quantum theory.
This completes the description of
this transition in the type IIA picture which
clearly has many loopholes and
some apparent inconsistencies.

\bigskip
\noindent
\subsection*{\it $M$ Theory Description of the Transition}

~~~~~Actually the description of this transition
in the $M$ theory picture is much
simpler and avoids all the problems mentioned.
The $M$ theory picture gives us one straight fivebrane
at $t=1$ spanning worldvolume
in the 012345 directions
and
another fivebrane with world
volume $\R^4\times \Sigma$ where $\R^4$ is the
world volume spanned by 0189 and
$\Sigma$ is a degenerate
Riemann surface in the 237 and 10 directions
which is described by the equation
$\sigma^nt=q$. The two fivebranes are connected by $k$ membranes.
In a vacuum configuration, the membranes are located at
the roots of $\sigma^n=q$.
Recall that $k$ of the $n$ roots are occupied by the membranes
and the remaining $n-k$ roots are not,
since one root cannot be occupied by two or more membranes
because $s$-configurations are not supersymmetric \cite{HW}.

\begin{figure}[htb]
\begin{center}
\epsfxsize=2.5in\leavevmode\epsfbox{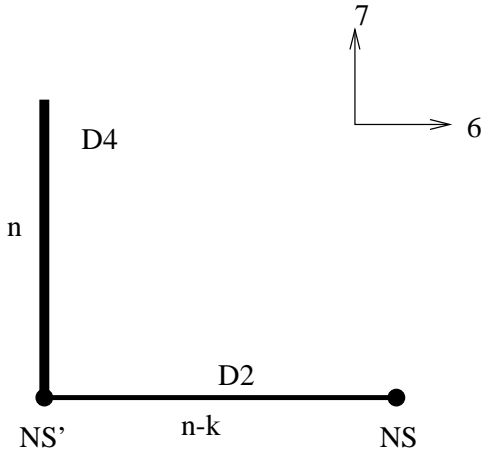}
\end{center}
\caption{Continuation past infinite Coupling for the Grassmannian Model.
There are $n$ upper D4 branes which end on NS$'$ brane and $n-k$ D2 branes 
stretched between the NS and NS$'$ branes. This is the type IIA description
of the transition which is done in the $M$ theory limit.}
\label{dual}
\end{figure}

In contrast to the situation in the type IIA picture,
the two fivebranes cannot avoid each other in space.
Thus the process of getting lower D4 branes
becomes unnecessary. Instead the two fivebrane
cross each other transversely in the space perpendicular
to the 01 directions.
In particular, they intersect at the $n$ roots of $\sigma^n=q$
at $t=1$, $x^{4,5,8,9}=0$.
Now the use of the transition of \cite{HW}
implies
\footnote{See also a discussion in \cite{BDG,DFK}.}
that, whenever
there was a membrane stretched before the transition,
there will be no membrane after that, and vice-versa.
That is, after the exchange in the $x^6$ direction
we are left with $n-k$
occupied positions of membranes and $k$
positions which are not occupied by
membranes. The resulting configuration
is sketched in figure \ref{dual} and 
the theory at hand thus describes $U(n-k)$ gauge theory with $n$
fundamental chiral multiplets.

The transition we have found implies
that there are two microscopically
inequivalent theories which are equivalent in the IR.
This is, in some sense, ``Seiberg's duality in two dimensions.''
What the brane picture demonstrates is how the transition past
infinite coupling implies this equivalence.

\bigskip
\noindent
\subsection*{\it Brane Proof of Level-Rank Duality}

~~~~~Let us recall that the number of vacua
for the $U(k)$ theory with $n$
fundamental fields is given by
the Witten index which is nothing but the
Euler characteristic of the Grassmannian $G(k,n)$.
It is given by $n\choose k$.
What we have just found is that the number
of vacua is consistent with the
transitions as the formula is invariant under the exchange
$k\leftrightarrow n-k$.
This is what is expected since
the Grassmannian $G(n-k,n)$ is ``dual'' to $G(k,n)$
and these are essentially the same.

There is one interesting point of view.
The transition we have found is nothing but the
level-rank duality of the WZW model,
as described in \cite{Schnitzer} (see also
\cite{tsu} and references
therein).
Recall that the dynamics of vacua of the Grassmannian sigma
model with target space $G(k,n)$ is described by the
$U(k)/U(k)$ gauged WZW model with the level of
$U(k)\sim SU(k)\times U(1)$ being $(n-k,n)$.
The level-rank duality says that
the space of conformal blocks of $SU(k)$ WZW model with level
$n-k$ is dual to the one of $SU(n-k)$ WZW model
with level $k$.
The $U(k)/U(k)$ gauged WZW model, being a topological field theory,
has as its correlation functions the dimension of
the space of conformal blocks. If we use this,
we see that our system is equivalent with the
$U(n-k)/U(n-k)$ gauged WZW model with the
level of $U(n-k)\sim SU(n-k)\times U(1)$
being $(k,n)$, which describes the sigma model with
target space $G(n-k,n)$ which in turn is given by the
$U(n-k)$ gauge theory with $n$ fundamental chiral multiplets. 
This is exactly what we have seen in the above discussion
of brane motions.
In other words, we have shown the level-rank 
duality in the brane framework.

It would be interesting to study
the interplay between this transition
and the methods used in the literature
to study this duality.
The brane picture demonstrates
that Seiberg Duality in four dimensions and
level rank duality in two dimensions follow
from the same qualitative level.

\subsection{More Brane Motion}

~~~~~We have performed all possible motions of branes in space for the $G(k,n)$
models.
We can now study transitions in other models by moving the branes around.
Let us add more branes into the picture.
Suppose that a lower D4 brane comes
from infinity in 23 to the origin.
We get a massless anti-fundamental field 
of $U(k)$. As in the last subsection,
the lower D4 brane can join an upper D4 brane
and move away from the NS$'$ brane in
the 456 directions. Let us assume that
the motion is only in the $x^6$ direction.
As discussed in section \ref{Class},
if the position of the D4 brane is to
the right of the NS$'$ brane,
the corresponding matter field decouples from
the low energy theory.
So we will discuss the case where the D4 brane is to the
left of the NS$'$ brane.
Then, there seems to be a flat
direction for which a D2 brane
can break and move in between the infinite
D4 brane and the NS$'$ - upper D4 branes system.
This is in contradiction with
Coleman's theorem which states that there are
no flat directions in two dimensions,
as mentioned in the last section.
To understand how this is possible
let us look at the $M$ theory solution for
this problem.

The equation which describes the addition of
the lower D4 brane in type IIA 
is
\beq
\sigma^nt=q\sigma.
\eeq
This demonstrates that indeed an infinite D4 brane in
type IIA
can decouple. In the $M$ theory picture two fivebranes are formed.
One at $\sigma=0$ and the other at
\beq
\sigma^{n-1}t=q.
\label{sinim}
\eeq
Next we want to break a membrane which will be stretched in
between the two fivebranes. However we encounter a problem since the two
equations have no common solution in the 23 direction.
That is unless $q$ is
equal to zero which is infinitely far away.
Such absence of solution may lead
us to conclude that the supersymmetric vacuum
is broken in contrast to the naive
expectation that there is a space of flat directions.

To understand better let us turn on twisted masses
$\tilm$ and $\btilm$ for the two fields
$Q$ and $\tilQ$, involved.
Then, the equation describing the fivebrane on the right is
\beq
\sigma^{n-1}(\sigma-\tilm)t=q(\sigma-\btilm).
\label{sinim2}
\eeq
First, consider the case with $\tilm=\btilm$. 
This gives a position $\tilm$ to the infinite D4 brane which in the
$M$ theory picture has the equation $\sigma-\tilm=0.$
Together with equation (\ref{sinim}) we see that the $x^7$ position of the
membrane is determined to be given by the equation
$t={q\over \tilm^{n-1}}$.
Thus, for a given $\tilm$, the $x^7$ and $x^{10}$ positions are fixed.
We also see that when the twisted mass is zero,
the membrane is running away
to infinity in this direction. There is still an option to turn on some
arbitrary values in the 89 directions
which will have the interpretation of
expectation value for meson fields.
This new flat direction corresponds to the
sigma model based on a complex one dimensional non-compact space
which is discussed in section 2.2.2.
The mechanism for freezing motion in this
direction is not clear.
We will assume that this motion is frozen.
The total set of vacua is thus $n-1$ massive
vacua corresponding to the solution
of (\ref{sinim}) and the vacua of this non-compact sigma model.

Second, we consider the generic values of $\tilm$ and $\btilm$.
As the equation (\ref{sinim2}) shows, the number of vacua is $n$.
Recall that a supersymmetric vacuum here
is interpreted as a stable brane
configuration which is in accord with the supersymmetry.
This condition is indeed satisfied.
For this case the Witten index, $\Tr(-1)^F$, does not depend on the value 
of the twisted mass parameter.

There are two generalizations to the models considered so far.
One type of generalization is to replace
the degenerate Riemann surface of the
NS$'$ brane and D4-branes system by a general Riemann surface.
The theory then will be of a D2 brane propagating
between a NS brane and the
Riemann surface. One possible system is to take a
Riemann surface which
describes a particular four dimensional theory like those studied in
\cite{WittenM5}. An example is discussed in the next section.

Another generalization is to replace
also the NS brane with an arbitrary Riemann
surface. In some sense the NS brane
is a very degenerate Riemann surface.
Then our aim would be to study the dynamics of
the D2 brane when it propagates
between two Riemann surfaces. More generally,
we can take a series of Riemann
surfaces localized at points in $x^6$. There can be arbitrary numbers of
D2 branes propagating between each two adjacent Riemann surfaces.

\subsection{D2 brane Propagation on $N=2$ Supersymmetric QCD.}

~~~~~We have studied in detail the propagation of
a D2 brane on a four dimensional
theory which is somewhat degenerate.
As a four dimensional theory it is 
weakly
coupled and frozen,
in the brane language. The closest realistic theory that 
the
$\CP^{n-1}$ model captures
in this brane realization is $N=1$ supersymmetric 
QCD.
The genus zero curve
which describes the chiral ring of the $\CP^{n-1}$ model
coincides with part of the genus zero curve which describes $N=1$ 
supersymmetric
QCD. The correspondence between the $\CP^{n-1}$ model and this theory was 
known
for a long time. This is reviewed in section 2.2.1.
The $\CP^{n-1}$ was used as a two dimensional toy model for
studying confinement and other four dimensional phenomena.
Both models share domain walls (solitons in the two
dimensional theory).
 From the brane point of view this is not a surprise.
The brane picture really provides
a string theory explanation for this
correspondence.
It tells us that studying the two dimensional model, probes
some qualitative features of the four dimensional theory.
At some cases, as for
the ratios between domain wall tensions versus soliton mass ratios, the
correspondence is even quantitative! This may be just the beginning of an
intersting interplay.

The aim of this section is to continue this approach and study
the propagation of the D2 brane on $N=2$ supersymmetric QCD in four 
dimensions.
This theory is described by
some D4 branes stretched between two NS branes.
We will choose,
as in the previous models, a NS brane at one end of the
D2 brane and in the other end we will take the 4d theory in question.
So let us start with a single D2 brane stretched between a NS brane to the 
left
and 2 NS$'$ branes to the right.
According to a conjecture by \cite{Kutasov},
this configuration describes
a $U(1)$ theory with adjoint field $x$ subject to
a superpotential $W=x^3$.
The $x$ field is associated with the 45 position of
the D2 brane.
This superpotential can be perturbed by moving the NS$'$ branes
in the 45 directions resulting in a superpotential which satisfies

\beq
{\partial W\over\partial x}=(x-a)(x-b).
\eeq
$a$ and $b$ are now the positions of the two NS$'$ branes in the 45 
directions.

Let us assume that this description
is correct and add more branes to the
picture. We put $n$ D4 branes
in the interval between the three NS branes
(the two NS$'$ branes are on the same $x^6$ position).
This process can be thought of as putting
the branes at far infinity in 45
directions and then slowly moving them to the origin of 45.
The resulting theory describes
a U(1) theory coupled to $n$ chiral fields of
charge +1 and their complex conjugates,
with the adjoint field subject to the
superpotential as above.

Next, let us move the NS$'$ branes
far apart in the $x^7$ direction. Such a
process does not change the superpotential.
The $x^7$ distance, being a real
parameter, can not enter the superpotential
if we assume, by supersymmetry,
holomorphic dependence.
We can now move the D4 branes in the $x^6$ direction
to touch the two NS$'$ branes. Once they touch, they can break and form a 
system
of $n$ finite D4 branes between two NS$'$ branes together with $n$ 
semi-infinite
lower (upper) D4 branes. To a four dimensional observer, living in the 0189
coordinates, this configuration is nothing but finite $N=2$ supersymmetric 
QCD
with gauge group $SU(n)$ coupled to $2n$ flavors represented by the
semi-infinite branes!

What is the two dimensional theory? We can repeat our analysis from the 
previous models and find that there are $n$ $Q_1$
and $\tilde Q_1$ fields coming from
localization near the first NS$'$ brane
and $n$ $Q_2$ and $\tilde Q_2$ fields
coming from localization
near the second NS$'$ brane. The adjoint field is 
still there subject to the superpotential.

Going back to the four dimensional theory,
varying the positions in the 456
directions is now interpreted as the Higgs
branch of this theory. In 
addition,
once the D4 branes break,
each part can move independently and we can get all
the models which are
derived from the finite theory by turning on expectation
values or mass terms.
In particular pure $N=2$ $SU(n)$ YM, and other models.
The Coulomb branch
of this theory is identified with the motion of the finite
D4 branes along
the NS$'$ branes in the 23 directions and is described by the
genus $n-1$ Riemann surface

\beq
R_n(\sigma)t^2+tP_n(\sigma)+Q_n(\sigma)=0,
\eeq
where $R,P$ and $Q$ are polynomials of order $n$ in $\sigma$.
For the two dimensional theory the 4d Higgs branch
is the space of complex mass
parameters together with $x^6$ and $A_7$ motions.
The 4d Coulomb branch is the
space of twisted mass parameters for the 2d theory.
We can continue with the identification further.
There is a complex modulus 
for the Riemann surface
which describes the coupling constant and theta angle of 
the four dimensional theory.
This is identified with the FI coupling and theta 
angle of the two dimensional theory.
This identification requires some explanation
which will be given below.
However it is sufficient to see that the FI coupling
of the two dimensional theory
receives non-perturbative corrections which 
come from contribution of
2d instantons which are stretched between the two NS$'$
branes. This is the first model in which such an effect happens for a two
dimensional FI parameter!
We find that two dimensional FI parameters behave 
just like four dimensional gauge couplings.

Let us go back to describe
more about the FI parameter. Consider a D2 brane
which propagates between
the Riemann surface of the 4d theory and the NS 
brane.
Let us assume that we have $m$ finite D4 branes and
$n_1 (n_2)$ upper (lower)
D4 branes. This describes $SU(m)$ gauge theory with $n_1+n_2$ flavors.
Near the first NS$'$ brane we have a $U(1)$
theory coupled to $m$ charge $+1$
fields $Q$ and $n_1$ charge $-1$ fields
$\tilde Q$. There is a FI term given by
the distance between the NS$'$ and the NS in the $x^7$ direction.
Near the second NS$'$ brane we have
a similar theory but it is a {\it different}
$U(1)$ coupled to $n_2$ charge $+1$ fields $Q$
and $m$ charge $-1$ fields $\tilde Q$.
The FI term is given by the $x^7$ distance between the second 
NS$'$
brane and the NS brane.
We see that there are two $U(1)$ theories with two
FI couplings. In the intermediate region
between the two NS$'$ branes both
$U(1)$ theories are broken and moving along
the four branes is a transition
between the two $U(1)$ theories.

There are two combinations for the FI couplings -- a sum and a
difference -- for which only the difference
is relevant for the physical 
system.
It is this quantity which is identified with
the $\tau$ parameter of the
Riemann surface for the 4d theory.

This situation is generalized easily to $k$ D2 branes
and also for including
more than two NS$'$ branes. It would be interesting
to study this system and 
the
2d -- 4d correspondence in more detail.

One more important point is in order.
At some special points
in the Coulomb branch of the four dimensional theory
there are additional massless multiplets -- monopoles or dyons.
If there are more
than one such additional massless multiplets then there can
be Higgs branches
which emanate from this singularity in the Coulomb branch.
Such branches are
not visible from a semi-classical point of view since the
singularities appear at strong coupling.
In terms of branes
what happens is that a Riemann surface becomes degenerate 
at
the singularity points and leads to disconnected Riemann surfaces.
This is interpreted as two fivebranes which locally come together and form a
massless hypermultiplet.
When the two branes move in a different direction 
away
from the singularity,
if this is possible, then a Higgs branch emanates from
the singular point.

The correspondence between the moduli space of vacua of
the four dimensional
theory and the space of parameters of
the two dimensional theory now has a 
new
prediction. Since the new Higgs branch which emanates is interpreted as 
complex
mass parameters for the two dimensional theory,
there are new couplings which
are not visible from the semi-classical
Lagrangian description. Clearly the 
new
couplings are visible from the brane point of view as the new direction for
which the two fivebranes can move.

\bigskip
\section{Solitons Via $M$ Theory}

~~~~~In this section, we give a description of the BPS saturated
solitons in $N=2$ field theory in two dimensions,
which were discussed in section 2,
in terms of branes in $M$ theory.

We recall from section 2.2 that the fundamental BPS soliton in the
$\CP^{n-1}$ model is the elementary field $Q$
in the fundamental representation of the flavor group
$SU(n)$ \cite{W79}.
The corresponding statement in the Type IIA description would be that
the fundamental soliton is the Type IIA string stretched between
the D2 brane and the D4 branes on the right.
In the $M$ theory description the string and the D2 brane
become a single membrane which winds around the eleventh
dimension in the region near the fivebrane on the right,
and is stretched between the two fivebranes.
We will show that this must be the case on topological grounds.
Recall also that the $\ell$-th soliton of the $\CP^{n-1}$
model (which interpolates the two vacua separated by $\ell$-steps)
is in the $\ell$-th anti-symmetric representation of $SU(n)$
($\ell$-th exterior product of the fundamental).
In the Type IIA description, this would be a bound state of
$\ell$ elementary strings stretched between the D2-brane
and the D4 branes.
We will also see this in $M$ theory including the fact
that they form the $\ell$-th anti-symmetric representation
of $SU(n)$,
as a consequence of a
constraint on the topology of the membrane which preserves
two of the four supersymmetries.
In addition, we compute the BPS mass by defining the
superpotential. These results are generalized to the case
with twisted masses
in which the soliton mass has not been computed from field theory
due to an ambiguity in defining the values of
the twisted superpotential. We will see that the soliton
masses can be determined unambiguously
in the brane framework.

Solitons in the supersymmetric $\CP^{n-1}$ model in two
dimensions are closely related to the domain
walls in the $N=1$ $SU(n)$ super Yang-Mills theory in four-dimensions,
as noted in section 2.2.
Recently, the domain wall separating the adjacent vacua
of super Yang-Mills theory
was studied in \cite{W97}
in the $M$ theory framework and claimed to be the D-branes for
QCD strings. It would be interesting to see the implication of the
present general description of the $\CP^{n-1}$ solitons
to the study of the domain walls.

\medskip
\subsection{The $\CP^{n-1}$ Solitons}

\bigskip
\subsection*{\it Brane Description of BPS Solitons}

~~~~~As explained in section 2, a soliton in two dimensional
field theory is a configuration of fields which
are at one vacuum in one spatial infinity
$x^1\to-\infty$
and are at another vacuum in the other spatial infinity
$x^1\to +\infty$.
Likewise, the soliton is described in $M$ theory
as a configuration of a membrane that depends on the spatial
coordinate $x^1$ so that it interpolates two different vacua
in this direction.

We first describe the solitons in the $\CP^{n-1}$ model.
Thus, we consider the configuration of section 4
with $k=1$, $n_1=n$ and $n_2=0$, where 
there are two fivebranes and one membrane stretched between them.
The two fivebranes are
a flat fivebrane at
$x^8=x^9=x^6=0,t=1$ and a curved fivebrane defined by
$\sigma^n t=q$ which is
at $x^4=x^5=x^6-L=0$.
For fixed $t$ and $q$ there are $n$ solutions for $\sigma$ which implies
that there are $n$ different vacua.
A vacuum configuration is given by a membrane,
with the time direction being omitted,
with world-volume a strip $\R\times I_j$ stretched between
the two fivebranes ($j=0,1,\ldots,n-1$),
where $\R$ is the one-dimensional space
with coordinate $x^1$ and
$I_j$ is a segment $0\leq x^6\leq L$
which is
located at $x^4=x^5=x^8=x^9=0$,
$t=1$ and  $\sigma=\e^{2\pi i j\over n}q^{1/n}$.
Therefore, a solitonic configuration will be given by
a membrane over a real two-dimensional surface
$\Sigma$ in the ten-dimensional space $\R^9\times S^1$
which is
stretched between the fivebranes
and interpolates two different segments, say $I_0$ and $I_{\ell}$,
in the $x^1$ direction.
Namely,
$\Sigma$ is a surface
with boundaries and two ends which are constrained by the following:
All the boundaries are in
the two fivebranes, and
at one end $x^1\to -\infty$, $\Sigma$ looks like $\R\times I_0$,
while at the other end
$x^1\to +\infty$, it looks like $\R\times I_{\ell}$.

Note here that what was the string and D2 branes in the Type IIA limit 
combine
into a single membrane which can only have boundaries on the fivebranes.
A string and a D2 brane become a single membrane on a Riemann surface just 
like
D4 branes and NS fivebranes become a single fivebrane in $M$ theory. 
This is indeed a supersymmetric configuration which breaks
locally one quarter of the supersymmetry charges.
Such a solitonic configuration is a BPS state
when $\Sigma$ is a supersymmetric cycle in the sense
that it preserves two of the four supersymmetries.
Here, we don't specify the precise condition for a cycle to be
supersymmetric in this situation,
although it should follow
from an argument as in \cite{BBS}.
It seems plausible, however, to require that a supersymmetric cycle
is a minimal surface.
Also, there must be two fermionic zero modes
coming from the $2=4-{4\over 2}$ broken supersymmetry.
Since some of the boundaries must be at $x^4=x^5=0$
and others at $x^8=x^9=0$, and since there is no other
condition involving $x^{4,5,8,9}$, the minimal surface
condition implies that $\Sigma$ is totally at $x^4=x^5=x^8=x^9=0$.
Thus, we can consider $\Sigma$ to be a surface in the
four-manifold with coordinates $t$ and $\sigma$.

In this paper, we do not touch the issue of existence and
uniqueness of the supersymmetric cycle.
Rather, we assume that there exists a unique BPS configuration
for each topological type, unless there is an obstruction
from fermionic zero modes.
The verification for this is an interesting open problem.

In what follows, we find a
restriction on the topology of such a cycle coming from
the boundary conditions.
It turns out that this restriction is
very strong and has a surprising consequence.
In particular, $\Sigma$ cannot be just a strip but has a topology of a
disc with holes, where each hole
is in the fivebrane on the right and
winds once around
the eleventh dimension.
This means that the configuration represents
a bound state of Type IIA strings
each of which carries a quantum number of the fundamental
representation of $SU(n)$.

\bigskip
\noindent
\subsection*{\it Fundamental Soliton $=$ Type IIA String}

~~~~~We first consider a solitonic configuration interpolating adjacent
vacua, i.e., $\sigma=1$, $t=1$ and
$\sigma=\e^{2\pi i\over n}$, $t=1$ (for this and the next part
of the section, we put $q=1$ for simplicity).
Since $\Sigma$ is stretched between the two fivebrane,
one boundary $J_l$
of $\Sigma$ is restricted to be in the
fivebrane on the left, i.e., at $t=1$ but $\sigma$ is free,
while
another boundary $J_r$
is restricted to be in the fivebrane on the right,
i.e., in the surface $\sigma^n=t^{-1}$. Due the condition at the two
ends, the boundary $J_l$ is a line in the
$\sigma$-plane at $t=1$
which connects the two points $\sigma=1$
and $\sigma=\e^{2\pi i\over n}$,
while the boundary $J_r$ is a line in the surface $\sigma^n=t^{-1}$
which connects the two points $\sigma=1$, $t=1$
and $\sigma=\e^{2\pi i\over n}$, $t=1$.
If we consider the projection to the $t^{-1}$-plane,
we see that $J_l$ is mapped to one point $t^{-1}=1$, but
$J_r$ is mapped to a circle starting and ending at the same point
which winds at least once around $t^{-1}=0$.
Here we choose a shortest path connecting $\sigma=1$ and
$\sigma=\e^{2\pi i\over n}$ so that the image of
$J_r$ in the $t^{-1}$-plane
winds exactly once counter-clockwise around $t^{-1}=0$.\footnote{
There are choices such that it winds $1\pm n,1\pm 2n,\ldots$
times, but we will see shortly that these cases are actually
equivalent to the case with winding number one.}
Thus, the surface has a circle boundary (of infinite length)
consisting of $J_l$, $J_r$ and two segments $I_0$ and
$I_1$ at $x^1=\mp\infty$,
which is mapped to a circle in the $t^{-1}$-plane
that winds once around $t^{-1}=0$.
Since $t^{-1}=0$ nor $t=0$ are not points in the
space-time, $\Sigma$ cannot be just a strip with disc topology
because, if it were a disc,
some point of $\Sigma$ would be mapped
either to $t^{-1}=0$
or to $t=0$.

\begin{figure}[htb]
\begin{center}
\epsfxsize=4.5in\leavevmode\epsfbox{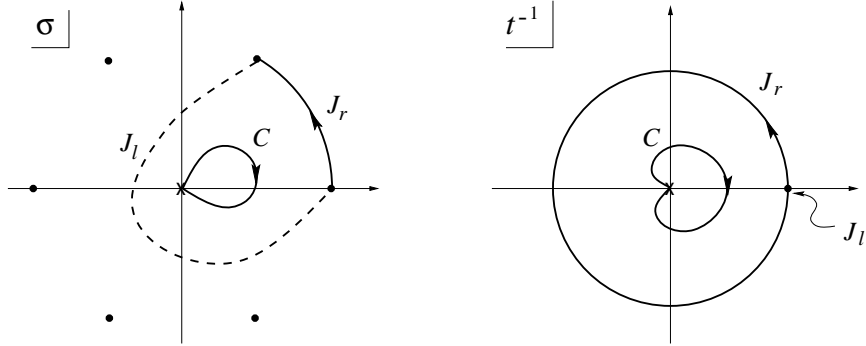}
\end{center}
\caption{The Image of $\Sigma$ in the $\sigma$- and the $t^{-1}$-Plane}
\label{ANN1}
\end{figure}

To avoid the points $t^{-1}=0$ and $t=0$,
$\Sigma$ must have another circle boundary $C$
that is mapped to a circle in the $t^{-1}$-plane which winds
once clockwise around $t^{-1}=0$.
The minimal choice of such a surface
is an annulus.
By the condition that all the boundaries of $\Sigma$ be in one of the
two fivebranes, the boundary $C$, not being at $t=1$,
must be in the fivebrane on the right.
In particular, $C$ winds once around
$t^{-1}=0$ in the $t^{-1}$-plane while satisfying the equation
$\sigma^n=t^{-1}$.
Then, the image of $C$ cannot be a circle which is totally
away from $t^{-1}=0$
since such a ``circle'' is mapped in the
$\sigma$-plane to a line starting and ending at two distinct
points related by the
$\e^{2\pi i\over n}$ rotation, which is not a circle.
This means that $C$ must start and end
at $\sigma=t^{-1}=0$. 
Thus, the image of the surface $\Sigma$ in the $\sigma$ and $t^{-1}$
planes look like the one depicted in Figure \ref{ANN1}.

\begin{figure}[htb]
\begin{center}
\epsfxsize=4.5in\leavevmode\epsfbox{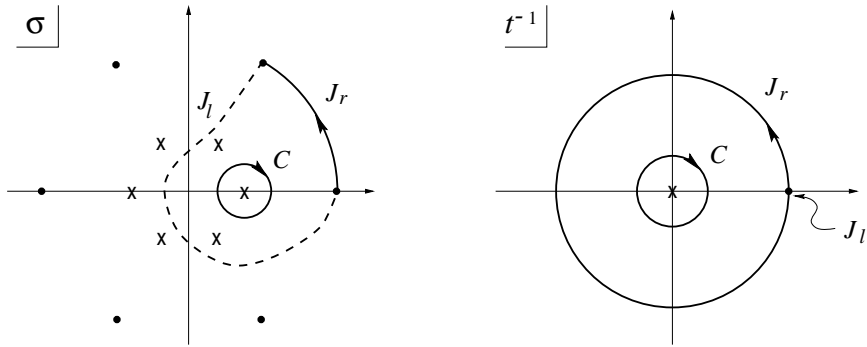}
\end{center}
\caption{One of the $n$ Possible Configurations}
\label{ANN}
\end{figure}

In the above description, however, it is not easy
to classify the topology of
circles starting and ending at $\sigma=t^{-1}=0$
since $\sigma=0$ is a degenerate zero.
In order to ``regularize'' this, we consider turning on small
distinct twisted masses for the fields $Q_i$. 
For convenience, we choose the masses to be
$\tilm,\tilm\e^{2\pi i\over n},\tilm\e^{4\pi i\over n},\ldots,
\tilm\e^{2\pi (n-1)i\over n}$, so that the fivebrane on the right is
described by
\beq
\sigma^n-\tilm^n=t^{-1}\,.
\label{reg}
\eeq
In this situation, the circle $C$ can wind around $t^{-1}=0$
without approaching $t^{-1}=0$, since the image in the $\sigma$-plane
can wind once around one
of the roots of $\sigma^n-\tilm^n=0$
because any root is a simple zero of this equation.
One such configuration looks like the one depicted
in Figure \ref{ANN}.
Since there are $n$ roots of $\sigma^n=\tilm^n$,
there are $n$-kinds of topological types of such configurations.

This can be interpreted as follows.
Recall that the roots $\tilm\e^{2\pi ij\over n}$ ($j=0,1,\ldots,n-1$)
at $x^6=L$
are interpreted as
the asymptotic position in $\sigma$ of the $n$ upper-half D4-branes
which are the parts of fivebrane wrapped once around the circle $S^1$
in the eleventh dimension. 
Recall also that the membrane wrapped once around the eleventh
dimension is interpreted as the Type IIA string.
Thus, the membrane configuration
in which the circle $C$ winds around the $j$-th root
can be considered
in the Type IIA string theory as a configuration in which
an open string is stretched between the D2-brane and
$j$-th upper-half D4-brane.
As explained in section 3,
such an open string generates the chiral multiplet $Q_j$.
In total, $Q_j$, $j=0,\ldots,n-1$ constitute the
fundamental representation of $SU(n)$.

In this way we have seen in $M$ theory description
that the solitons
interpolating adjacent vacua are interpreted as
the elementary Type IIA strings which in turn give rise to
the elementary chiral multiplet $Q$ in the fundamental
representation of the flavor group $SU(n)$,
in accord with the field theory knowledge \cite{W79}
(see section 2.2).

\bigskip
\noindent
\subsection*{\it The Exclusion Principle of Type IIA Strings}

~~~~~Let us next consider more general solitons, interpolating
$I_0$ and $I_{\ell}$ for $\ell=2,3,\ldots n-1$,
where we work in the ``regularized'' configuration (\ref{reg})
in which $I_0$ is at $\sigma=(1+\tilm^n)^{1/n}$, $t=1$ while
$I_{\ell}$ is at $\sigma=(1+\tilm^n)^{1/n}e^{2\pi i\ell\over n}$.
In this case, the boundary $J_l$ at the fivebrane on the left
is a line in the $\sigma$-plane at $t=1$ connecting
$I_0$ and $I_{\ell}$
while the boundary $J_r$ at the fivebrane on the right
is a line in the surface $\sigma^n-\tilm^n=t^{-1}$
connecting $I_0$ and $I_{\ell}$. In particular,
the image of $J_l$ in the $t^{-1}$-plane is at a point $t^{-1}=1$
while the image of $J_r$ starts and ends at that point
winding $\ell$-times around $t^{-1}=0$. 
Thus, the surface $\Sigma$ has a large boundary circle (which is
actually of infinite length) consisting of $J_l$, $J_r$ and the two
ends, which winds $\ell$-times counter-clockwise around
$0$ in the $t^{-1}$-plane.

To avoid the points $t^{-1}=0,\infty$ which are not in the space-time,
$\Sigma$ must have some other boundary circles.
Since the large boundary winds $\ell$-times around $0$
in the $t^{-1}$-plane, there must be other $\ell$ boundary circles
$C_1,\ldots,C_{\ell}$
(which can be connected and rejoined), each of which
winds once clockwise around $0$ in the $t^{-1}$-plane.
Since these boundaries
must be in the fivebranes and since these are not
identically
at $t^{-1}=1$, they all must be in the fivebrane on the right, namely,
at $x^6=L$, $\sigma^n-\tilm^n=t^{-1}$.
The fact that $C_i$ are circles means that each of them
winds once around one of the $n$ roots of $\sigma^n-\tilm^n=0$
in the $\sigma$-plane.

How these boundary circles $C_i$ choose the roots?
We show that two or more
distinct circles cannot choose one common root.
In other words, one boundary cannot wind twice or more times around
one root.
We show this in the case $\ell=2$ which captures the essence.

\begin{figure}[htb]
\begin{center}
\epsfxsize=4.5in\leavevmode\epsfbox{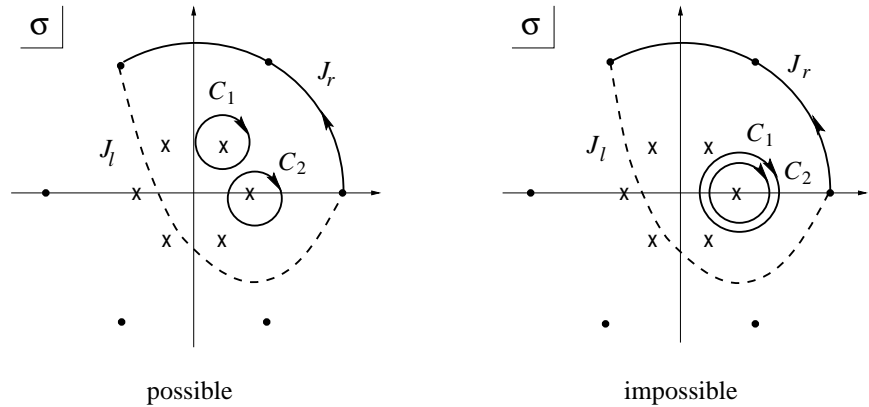}
\end{center}
\caption{Possible and Impossible Configurations with $\ell=2$}
\label{PI}
\end{figure}
Suppose that the two circles $C_1$ and $C_2$
winds a common root, say $\sigma=\tilm$, as in the RHS of
Figure \ref{PI}.
We first compactify the surface $\Sigma$ by
capping 
the three boundary circles by three discs.
The three boundaries are $C_1$, $C_2$ and
the large boundary circle (of infinite length) consisting
of $J_l$ and $J_r$ joined by $I_0$ and $I_{\ell}$.
We denote the compactified surface by $\overline{\Sigma}$.
In the case where $C_1$ and $C_2$ are rejoined, we consider
the corresponding discs to be rejoined as well.
The projection of $\Sigma$ to the
$\sigma$-plane can be considered as a complex valued function.
Then, one can deform it so that it defines a meromorphic
function of the compact Riemann surface $\overline{\Sigma}$
with respect to a suitably chosen complex structure.
By the boundary condition, $\sigma-\tilm$ is a function
which has one simple pole and two simple zeros (or one double zero).
The Riemann-Roch theorem implies that there is no Riemann surface
having a meromorphic function with this property.
This completes the proof that the RHS of Figure \ref{PI}
is impossible. Note that the LHS is possible
because it implies that $\sigma-\tilm$ has one simple zero and one
simple pole, which is possible for $\overline{\Sigma}=\CP^1$.

\begin{figure}[htb]
\begin{center}
\epsfxsize=1.9in\leavevmode\epsfbox{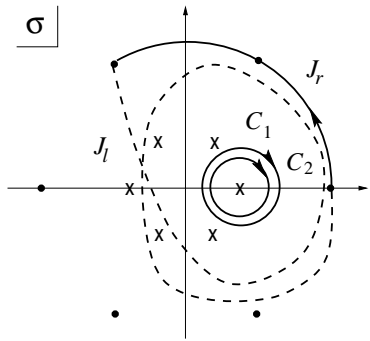}
\end{center}
\caption{Possible but Non-BPS Configuration}
\label{nonSUSY}
\end{figure}

If the boundary on the left, $J_l$, winds once
around $\sigma=\tilm$ as in Figure \ref{nonSUSY},
two distinct small boundaries can wind around $\sigma=\tilm$
on topological grounds.
However, this configuration has twice as many fermionic zero modes as
the one in Figure \ref{ANN} or in the LHS of Figure \ref{PI}.
Since the latter, being a BPS configuration, preserves half of the four
supersymmetries, it carries two fermionic zero modes.
Therefore, the configuration in Figure \ref{nonSUSY}
has four zero modes and it is likely that these can be interpreted
as the Goldstino associated with the breaking of
all of the supersymmetry.
Namely, we claim that it is not a BPS configuration.

\begin{figure}[htb]
\begin{center}
\epsfxsize=4.5in\leavevmode\epsfbox{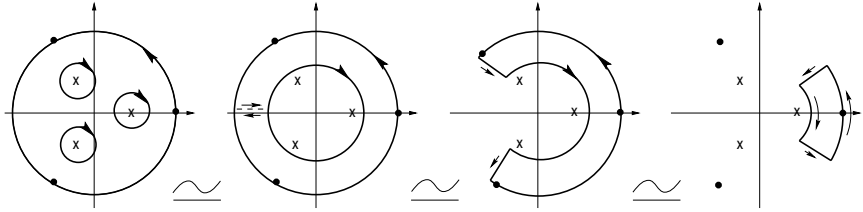}
\end{center}
\caption{The Process of Shrinking in the Case $\ell=n$
(The $\sigma$-Planes)}
\label{SH}
\end{figure}

Let us consider the case $\ell=n$, where the large boundary
winds $t^{-1}=0$ $n$-times. By the ``exclusion principle''
of the boundaries
which we have just proved, there are $n$-circle boundaries
$C_1,\ldots,C_n$ where each $C_j$ 
winds once around each $\tilm\e^{2\pi ij\over n}$
of the roots of $\sigma^n=\tilm^n$. This configuration is actually
unstable as it can be shrunk to a point, as the Figure \ref{SH}
shows.
As a consequence of this, we see that the solitonic configuration
interpolating $I_0$ and $I_{\ell+n}$ in the counter-clockwise
direction in the $\sigma$-plane
is equivalent to a configuration interpolating
$I_0$ and $I_{\ell}$ in the same direction, where $m$ is any positive
integer.

\begin{figure}[htb]
\begin{center}
\epsfxsize=3.5in\leavevmode\epsfbox{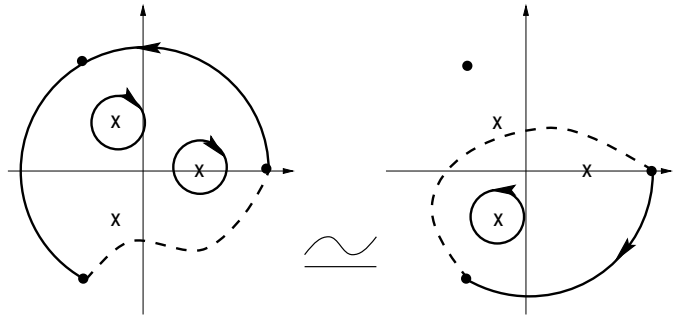}
\end{center}
\caption{$\ell=n-1$ is equivalent to $\ell=-1$}
\label{Anti}
\end{figure}

By a similar reasoning, one can show that the solitonic
configuration interpolating $I_0$ and $I_{n-1}$ in the
counter-clockwise direction is equivalent to the
solitonic configuration interpolating $I_0$ and
$I_{-1}$ in the clockwise direction. (See Figure \ref{Anti})
Note that the orientation of the small boundary circle
on the RHS of the figure is inverse to the one we have been
considering.

All these have quite natural interpretations.
First of all, a configuration of membrane
with several
small boundaries which are attached to the fivebrane on the
right and wind once around the circle $S^1$ in the eleventh direction
is interpreted in Type IIA string theory as some
bound state of elementary open strings and a D2-brane,
where the string end
points carry quantum numbers of $SU(n)$ fundamental representation.
For a solitonic configuration interpolating $I_0$ and $I_{\ell}$,
there are $\ell$ such small boundaries, meaning that there are $\ell$
string end points, and thus, the corresponding
state is in some $\ell$-fold tensor product of the fundamental
representation of $SU(n)$. The ``exclusion principle''
of the small boundaries of a
BPS configuration implies that this tensor product
representation for the BPS states
is actually the anti-symmetric tensor representation,
i.e., $\ell$-th exterior power
\beq
\bigwedge^{\ell}\C^n
\eeq
of the fundamental representation $\C^n$. This representation has
dimension
\beq
{n\choose \ell}
\eeq
which coincides with the number of possible topological types.
The fact that configuration interpolating $I_0$ and $I_{n-1}$
is equivalent with the one interpolating $I_0$ and $I_{-1}$,
with the orientation of the small boundary circles being flipped,
corresponds to the equivalence
\beq
\bigwedge^{n-1}\C^n\cong \overline{\C^n}
\eeq
as a representation of $SU(n)$.
That there is no stable configuration interpolating $I_0$ and $I_n$
means that there are no BPS states in such a sector.

These reproduce
what we know from field theory argument \cite{W79,KK,AL,CV2}
as essential properties of the $\CP^{n-1}$ solitons
in a very interesting way.

\medskip
\bigskip
\subsection{The Twisted Superpotential and The BPS Mass}

\medskip
~~~~~Finally, we compute the mass of these solitons.
As explained in section 2.2, the mass of a BPS state is
given by the difference of the values of superpotential
at the two spatial infinities. Thus, we start by defining
superpotential in the $M$ theory framework, by essentially following
the path made in \cite{W97}.
Note that what we call ``superpotential'' here is actually
twisted superpotential since we are considering a theory
in which $\UV$ is unbroken.

Configuration of a membrane at a fixed two-dimensional
space-time point is a segment $I$
in the nine-dimensional space
with coordinates $x^2,\ldots,x^{10}$
which is stretched between the two fivebranes.
A superpotential is a function of the space of such
configurations $I$ satisfying the two basic properties:
It is a holomorphic function, and its critical points
are vacuum configurations in which the supersymmetry
is totally unbroken.
As we have seen in section 4, a vacuum configuration is 
a straight segment in the $x^6$ direction, namely,
one of the $n$ segments $I_0,I_1,\ldots, I_{n-1}$ parametrized by
$0\leq x^6\leq L$.
In order for the ``holomorphic function'' to make sense,
we need to introduce the complex structure of the space of
configurations. Since we are interested in the configurations
which are totally at $x^4=x^5=x^8=x^9=0$ and in $0\leq x^6\leq L$,
we may consider a configuration $I$ as given by a pair of functions
$\sigma(x^6)$, $t(x^6)$ of the segment $0\leq x^6\leq L$
which are constrained by the condition
$t(0)=1$ and  $\sigma(L)^n-\tilm^n=t(L)^{-1}$
for ending on the fivebranes.
\footnote{Of course, the restriction to $x^4=x^5=x^8=x^9=0$,
$0\leq x^6\leq L$
is not essential and we could develop the following argument for
general configuration, but it makes no difference
in the final result.}
Then, the superpotential must be a holomorphic functional
on the space of such pair of functions.

As in \cite{W97}, at least locally
the superpotential can be defined in the following way
up to additive constant.
Let $\Omega$ be the holomorphic two form
\beq
\Omega=\dd\sigma\wedge{\dd t\over t}.
\eeq
Given two configurations, say $I$ and $I^{\prime}$,
the difference of the values of superpotential is defined by
\beq
\tilW(I)-\tilW(I^{\prime})=\int_{\Sigma}\Omega
\label{defsup}
\eeq
where $\Sigma$ is a one-parameter family of configurations 
interpolating $I$ and $I^{\prime}$.
Then, it is easy to see that
this satisfies the basic requirements.
Consider a variation of a configuration
$I\to I+\delta I$. Then,
\beq
\delta \tilW=\int_I
\left(\delta\sigma{\dd t\over t}-{\delta t\over t}\dd\sigma\right).
\eeq
The fact that this is independent on
$\delta\overline{\sigma}$ and $\delta\overline{t}$
means that $\tilW$ indeed depends holomorphically on $I$.
Also, a critical
configuration is given by the one satisfying
$\dd t=\dd\sigma=0$, i.e., a straight segment in which
$t(x^6)$ and $\sigma(x^6)$ are constant functions.

One may wonder how to fix the normalization of the
superpotential.
Note that the holomorphic two form $\Omega$ is related to
the one ${\bf \Omega}$ given in (\ref{Omst})
associated with the eleven-dimensional
space-time metric
(\ref{elevenmet}) by ${\bf \Omega}=\elel^{\,3}\Omega$ where
$\elel$ is the eleven-dimensional Planck length.
Later we will see that the above normalization of $\tilW$ is correct
up to a numerical factor.

Now we need to check that a superpotential is actually defined globally
by (\ref{defsup}) up to additive constant.
First of all, as we have observed in the above discussion,
two configurations cannot always be
interpolated by a one parameter family of segments.
Thus, we must relax the condition on $\Sigma$ by
allowing it to have some boundary circles which are
in the fivebranes.
Namely, $\Sigma$ is a real two-dimensional
surface in the four-manifold
with complex coordinate $t$ and $\sigma$ which has boundaries
like
\beq
\partial \Sigma=I-I^{\prime}+J_R-J_L
+C_1+\cdots +C_s
\eeq
where $J_{L,R}$ are lines in the left and right fivebranes
connecting the end points of $I$ and $I^{\prime}$,
and $C_1,\ldots,C_s$ are circles in the fivebrane on
the right.
What we must show is that the difference $\tilW(I)-\tilW(I^{\prime})$
given by (\ref{defsup})
is independent on the choice of such a surface $\Sigma$.
Let us take another surface $\Sigma^{\prime}$ with the boundary
\beq
\partial \Sigma^{\prime}
=I-I^{\prime}+J_R^{\prime}-J_L^{\prime}
+C_1^{\prime}+\cdots +C_{s^{\prime}}^{\prime}.
\eeq
Then, the difference of the superpotential changes by
\beqa
\int_{\Sigma-\Sigma^{\prime}}\Omega&=&
\int_{\partial \Sigma-\partial\Sigma^{\prime}}\sigma{\dd t\over t}\\
&=&\left(\oint_{J_R-J_R^{\prime}}-\oint_{J_L-J_L^{\prime}}
+\sum_{j=1}^s \oint_{C_j}
-\sum_{j^{\prime}=1}^{s^{\prime}}\oint_{C_{j^{\prime}}^{\prime}}\right)
\sigma{\dd t\over t}
\eeqa
where we have used $\Omega=\dd(\sigma \dd t/t)$.
Note first that the integration of $\sigma\dd t/t$ over
$J_L-J_L^{\prime}$ vanishes, since $t$ is constant $t\equiv 1$
along $J_L$ and $J_L^{\prime}$ which are in the fivebrane on the left.
Other boundaries are all closed circles
in the fivebrane on the right in which
\beq
\sigma{\dd t\over t}=
-n{\sigma^n\dd \sigma\over \sigma^n-\tilm^n}
=-n\,\dd\sigma-\tilm^n{\dd\sigma\over \sigma^n-\tilm^n}.
\label{diff}
\eeq
Thus, the difference $\tilW(I)-\tilW(I^{\prime})$
changes by the sum of residues
of the differential of the second term
on the right hand side, which are proportional to $\tilm$.
This vanishes in the limit $\tilm\to 0$.
Therefore, in this $\tilm=0$ case,
the superpotential is indeed globally
defined by (\ref{defsup})
up to additive constant.

The actual computation of the superpotential is straightforward.
Since we are interested in the mass of the BPS solitons,
we compute the difference of the values at the vacuum configurations
$I_0$ and $I_{\ell}$.
Then, we can take as $\Sigma$ the solitonic membrane configuration
which we discussed in the previous part of this section.
Since $\Omega=\dd(\sigma\dd t/t)$, we have
\beqa
\tilW(I_{\ell})-\tilW(I_0)&=&
\int_{\partial \Sigma}\sigma{\dd t\over t}\\
&=&\left(
\int_{I_{\ell}}-\int_{I_0}+\int_{J_r}-\int_{J_l}
+\sum_{j=1}^{\ell}\oint_{C_j}\right)
\sigma{\dd t\over t}.
\eeqa
Since $I_{\ell},I_0,J_l$ are at $t=1$, the corresponding integrals
vanish. Also, the integration over the small boundaries $C_j$
in the fivebrane on the right vanish
in the limit $\tilm\to 0$ as we have seen using (\ref{diff}).
Thus, the only non-vanishing term is the integration over the path
$J_r$ in the fivebrane on the right in which
$$
\sigma{\dd t\over t}\,=\,-n\,\dd\sigma
$$
in the case $\tilm=0$ by (\ref{diff}).
Since $J_r$ is a path connecting
$\sigma=q^{1/n}$ and $\sigma=q^{1/n}\e^{2\pi i\ell\over n}$,
the difference of the superpotential values is
\beqa
\tilW(I_{\ell})-\tilW(I_0)&=&-\int_{J_r}\,n\,\dd\sigma\nonumber\\
&=&n\,q^{1/n}\left(\,1-\e^{2\pi i\ell\over n}\,\right).
\label{delW}
\eeqa

\medskip
The mass of the BPS soliton is basically the absolute value
of this difference (\ref{delW}).
Here we comment on the reason for this, which shows also that
the normalization of the definition of $\tilW$
is correct up to a numerical factor (i.e. a factor which is
independent of the parameters of the system).
The membrane action contains the volume of its worldvolume, and hence,
the energy is the area of its spatial part.
Even though the area is generally infinite,
we may regularize it by
considering a difference of the area of
an exited configuration and the one of
the gound configuration.
It is a natural guess, although we do not presently
have a proof of it,
that such an area is bounded from below
by the absolute value of
the integration of the holomorphic two form ${\bf \Omega}$
of the $x^{2,3,7,10}$ part of the space-time (\ref{Omst}),
and that the BPS configuration with two unbroken supersymmetry
saturates this.
For a finite time interval ${\sl \Delta} x^0$,
the action is given by
\beq
E{\sl \Delta} x^0={1\over \elel^{\,3}}\int
\dd x^0\left|\int_{\Sigma}{\bf\Omega}\,\right|
={1\over \elel^{\,3}}{\sl \Delta} x^0
\left|\int_{\Sigma}\elel^{\,3}\,\Omega\,\right|
={\sl \Delta} x^0\left|\int_{\Sigma}\Omega\,\right|\,.
\eeq
This is the reason why the absolute value of
(\ref{delW}) is the mass of the BPS soliton
(up to a numerical factor which we have not been
careful enough to fix).
In view of the identification of the parameters
(\ref{iden}), the mass of the soliton is given by
\beq
M_{\ell}\,=\,n\Lambda\left|\,1-\e^{2\pi i\ell/n}\,\right|\,,
\eeq
which coincides with what we know from field theory
(\ref{solitonmass})
\cite{KK,AL,CV2,CV}, up to a numerical factor.

\bigskip
\noindent
\subsection*{\it General Twisted Mass}

\begin{figure}[htb]
\begin{center}
\epsfxsize=2.5in\leavevmode\epsfbox{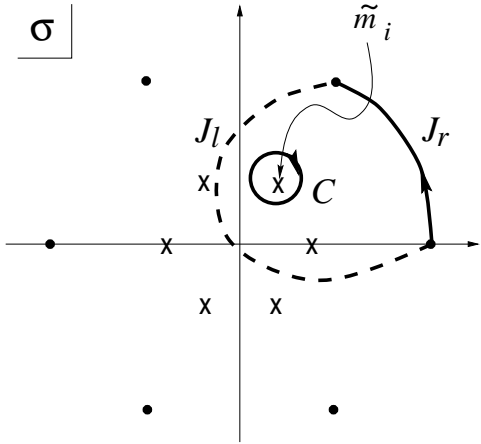}
\end{center}
\caption{The Contours}
\label{twis}
\end{figure}

~~~~~It is easy to extend the above analysis to the case
where general twisted masses $\tilm_i$ are turned on.
The one form $\sigma\dd t/t$ is expressed
in the fivebrane on the right as
\beq
\sigma{\dd t\over t}=
-\sigma\sum_{i=1}^n{\dd\sigma\over\sigma-\tilm_i}
=-n\dd\sigma-\sum_{i=1}^n{\tilm_i\dd\sigma\over \sigma-\tilm_i}
\,.
\label{oneformtwis}
\eeq
Thus,
there is an ambiguity in defining the superpotential
due to the residue of this form, as we have seen right above
in a special case. Namely, the ambiguity
is proportional to $2\pi i\tilm_i$. This is exactly the ambiguity
we have observed in section 2.2 in the field theory discussion
(up to the usual factor of $2\pi$).
However, we can nevertheless define unambiguously
the central charge, or the masses of the BPS solitons.
For a BPS configuration given by a surface $\Sigma$,
the central charge is simply defined as the integration of
$\Omega$ over $\Sigma$:
\beq
\widetilde{Z}=\int_{\Sigma}\Omega\,.
\label{Ztwis}
\eeq
Consider, for example, a configuration interpolating neighboring
vacua with a single small boundary circle $C$ which winds around
$\sigma=\tilm_i$, as depicted in Figure \ref{twis}.
Then, the integration (\ref{Ztwis})
is reduced to the contour integration of the one form
(\ref{oneformtwis}) along the solid lines $C$ and $J_r$.
This is actually the same as the integration over the path
$P_i$ considered in section 2.2, Figure \ref{ambiguity},
since $J_r+C-P_i$
is a boundary of a surface on which the one form
(\ref{oneformtwis}) has no residue.
Namely, the same central charge can be expressed in two ways
\beq
\widetilde{Z}\,=\,\int_{P_i}\sigma{\dd t\over t}
\,=\,\int_{P_0}\sigma{\dd t\over t}+2\pi i\tilm_i\,,
\eeq
where $P_0$ is the notation for $J_r$ used in section 2.2
(compare Figure \ref{twis} and Figure \ref{ambiguity}).
This corresponds to the equation (\ref{eqambi}).
The first expression is interpreted as the difference of the values
of the superpotential associated with a choice of the path ($P_i$)
and the second is interpreted as the sum of
the one associated with another choice ($P_0$)
and the twisted mass times the $U(1)$ charge carried by the soliton.
In general, the central charge is expressed as
\beq
\widetilde{Z}\,=\,\Delta \tilW\,+\,2\pi i\sum_{i=1}^n\tilm_iS_i\,,
\eeq
where $S_i$ is the charge of the $i$-th $U(1)$ of the
group $U(1)^n$, the subgroup of the flavor group
$U(n)$ (modulo the gauge group $U(1)$) which remains
unbroken by the twisted masses.
Note that a change of the path defining the superpotential
changes the $U(1)$ charges
$S_i$ by an amount related to the topological charge.

Finally, let us see what happens if we send one of the mass,
say $\tilm_n$, to infinity. If we do this by keeping fixed
$\Lambda_L$ and $\e^{i\theta_L}$
defined by
\beq
\Lambda^n\e^{i\theta}=\tilm_n\Lambda_L^{n-1}\e^{i\theta_L}\,,
\eeq
then, one of the $n$ roots of
$\prod_{i=1}^n(\sigma-\tilm_i)=q=\Lambda^n\e^{i\theta}$
is of order $\tilm_n$
and goes to infinity, while the rest becomes
the $n-1$ roots of
$\prod_{i=1}^{n-1}(\sigma-\tilm_i)=\Lambda_L^{n-1}\e^{i\theta_L}$
and are finite.
Namely, one of the $n$ vacua runs away to infinity and
only $n-1$ of them remain.
The mass of the BPS soliton interpolating two vacua
will stay finite if the small boundary circles do not
wind around $\sigma=\tilm_n$ and the boundary $J_r$ stays finite.
However even a BPS soliton may disappear
by acquiring an infinite mass, if the boundary circle winds around
$\sigma=\tilm_n$ or the boundary $J_r$ is infinitely elongated.
If we consider the process
\beq
\tilm=0\longrightarrow\tilm={\rm diag}(0,0,\ldots,0,\tilm_n=\infty),
\eeq
then, the soliton spectrum changes as
\beqa
&&\bigwedge^{\ell}\C^n\longrightarrow\bigwedge^{\ell}\C^{n-1}\,,
\qquad \ell=1,\ldots,n-2\\
&&\bigwedge^{n-1}\C^n\longrightarrow \emptyset\,.
\eeqa

\vfill
\noindent
{\Large\bf Acknowledgements}

\medskip
It is a pleasure to thank discussions with Ofer Aharony, John Brodie,
Jan de Boer, Kenneth Intriligator, Andreas Karch, Hitoshi Murayama,
Hirosi Ooguri, Yaron Oz, Matthew Strassler, Cumrun Vafa,
Nicholas Warner, Edward Witten, Zheng Yin
and Alberto Zaffaroni.
A.H. would like to thank Aspen Center for Physics
for hospitality while this work was in final stages.
K.H. would like to thank
the Institute for Advanced Study where this work was initiated and Rutgers
Physics Department, for hospitality.
The research of A.H. is supported in part by NSF grant PHY-9513835.
The research of K.H. is supported in part by NSF grant PHY-9514797
and DOE grant DE-AC03-76SF00098.

\end{document}